\documentclass[12pt]{article}

\usepackage{graphicx}
\usepackage{newtxtext}
\usepackage{newtxmath}
\usepackage{natbib}
\usepackage{hyperref}
\hypersetup{
    colorlinks = true,
    urlcolor   = blue,
    citecolor  = black,
}

\newcommand{\RomanNumeralCaps}[1]
\linenumbers
\usepackage{tikz}
\usetikzlibrary{decorations.markings}
\usepackage{mathrsfs}
\usepackage[all,cmtip]{xy}
\usepackage{mleftright}
\usepackage{booktabs}
\usepackage{enumitem}
\usepackage[outdir=./]{epstopdf}
\usepackage{caption}
\usepackage{subcaption}
\usepackage{placeins}
\usepackage{comment}

\newtheorem{thm}{Theorem}[section]

\newtheorem{problem}[thm]{Problem}
\numberwithin{equation}{section}

\newcommand{\IGNORE}[1]{}
\newcommand{\ignore}[1]{}

\newcommand{\mbb}[1]{\mathbb{#1}}

\newcommand{\mc}[1]{\mathcal{#1}}

\newcommand{\jt}{\textstyle}

\newcommand{\lla}{\left\langle}
\newcommand{\rra}{\right\rangle}

\newcommand{\la}{\langle}
\newcommand{\ra}{\rangle}
\newcommand{\disc}{{\text{disc}}}
\newcommand{\ess}{{\text{ess}}}

\newcommand{\PP}{\operatorname{P}}
\newcommand{\PR}{\operatorname{R}}
\DeclareMathOperator*{\argmax}{argmax}

\newcommand{\bds}{\boldsymbol}

\newcommand{\tol}{\text{tol}}

\def\bnabla{\boldsymbol{\nabla}}
\def\a{\boldsymbol{a}}
\def\u{\boldsymbol{u}}

\def\x{\boldsymbol{x}}

\def\bxi{\boldsymbol{\xi}}

\def\bxi{\boldsymbol{\xi}}
\def\RR{\mathbb{R}}
\def\TT{\mathbb{T}}
\def\CC{\mathbb{C}}
\def\ZZ{\mathbb{Z}}

\def\bL{\mathbf{L}}

\def\wopt{\widetilde{w}}

\def\sigp{\sigma_{\text{p}}}
\def\sigess{\sigma_{\text{ess}}}

\def\mumax{\mu_{\text{max}}}

\renewcommand{\L}{\mathcal L}

\def\A{\mathcal{A}}
\def\C{\mathcal{C}}
\def\M{\mathcal{M}}
\def\O{\mathcal{O}}

\def\W{\mathcal{W}}

\begin{document}
\title{On the inviscid instability of \\ the 2D Taylor-Green vortex}

\author{Xinyu Zhao$^{1,2}$,   
Bartosz Protas$^{1,}$\thanks{Email address for correspondence: bprotas@mcmaster.ca} \ and Roman Shvydkoy$^3$
\\ \\ 
$^1$Department of Mathematics and Statistics, McMaster University \\
Hamilton, Ontario, L8S 4K1, Canada \\ 
\\
$^2$Department of Mathematical Sciences, New Jersey Institute of Technology, \\
Newark, NJ 07103, USA \\
\\
$^3$Department of Mathematics, Statistics and Computer Science, \\
University of Illinois at Chicago, Chicago, IL 60607, USA}

\date{\today}
\maketitle

\begin{abstract}
   We consider Euler flows on two-dimensional (2D) periodic domain and
  are interested in the stability, both linear and nonlinear, of a
  simple equilibrium given by the 2D Taylor-Green vortex. As the first
  main result, numerical evidence is provided for the fact that such
  flows possess unstable eigenvalues embedded in the band of the
  essential spectrum of the linearized operator. However, the unstable
  eigenfunction is {discontinuous at the hyperbolic stagnation
    points of the base flow and its regularity is consistent
    with the prediction of \citet{Lin2004}. This eigenfunction gives
    rise to an exponential transient growth with the rate given by the
    real part of the eigenvalue followed by passage to a nonlinear
    instability.}  As the second main result, we illustrate a
  fundamentally different, non-modal, growth mechanism involving a
  continuous family of uncorrelated functions, instead of an
  eigenfunction of the linearized operator.  Constructed by solving a
  suitable PDE optimization problem, the resulting flows saturate the
  known estimates on the growth of the semigroup related to the
  essential spectrum of the linearized Euler operator as the numerical
  resolution is refined. These findings are contrasted with the
  results of earlier studies of a similar problem conducted in a
  slightly viscous setting where only the modal growth of
  instabilities was observed. This highlights the special stability
  properties of equilibria in inviscid flows.
\end{abstract}

\begin{flushleft}
Keywords: Inviscid flows; Instability; Essential spectrum;  Variational methods;
\end{flushleft}



\section{Introduction}

We study the stability of a two-dimensional (2D) flow of an
incompressible ideal fluid described by the classical Euler system
subject to periodic boundary conditions
\begin{subequations}\label{eq:euler}
\begin{align}
{\partial_t \bds u} + (\bds u \cdot \nabla)\bds u &= - \nabla p, &\qquad &(\bds x, t) \in 
 \mbb T^2 \times (0, +\infty), \label{eq:eulerA} \\
\nabla \cdot \bds u &= 0, &\qquad &(\bds x, t) \in \mbb T^2 \times (0, +\infty), \\
\bds u (\bds x, 0)&= \bds u_0(\x),  &\qquad  &\bds x\in \mbb T^2,  \label{eq:eulerC} 
 \end{align}
\end{subequations}
where $\mbb T^2 := \mbb R^2/ (2 \pi\mbb Z)^2$ (``$:=$'' means ``equal
to by definition''), whereas $\x = (x_1,x_2)^T$ and $t \ge 0$ are,
respectively, the spatial coordinate and time. In \eqref{eq:euler},
$\bds u = \u(t, \x) = (u_1, u_2)^T$ is the velocity field,
$p = p (t, \bds x)$ the scalar pressure, whereas $\bds u_0$ is the
initial condition for the velocity field, assumed divergence-free,
$\nabla \cdot \bds u_0 = 0$.

Computing the curl of both sides of {\eqref{eq:eulerA}}, the equation for
the scalar vorticity ${\omega := \nabla^\perp \cdot \bds u}$, where
${\nabla^\perp := \left(-\partial_{x_2}, \partial_{x_1}\right)}$, is
\begin{subequations}
\label{eq:vort:euler}
\begin{align}
\partial_t \omega + (\bds u \cdot \nabla)\omega = 0&, &\qquad &(\bds x, t) \in 
 \mbb T^2 \times (0, +\infty), \\
\bds u = {\nabla^\perp \Delta^{-1} \omega}&,  &\qquad &(\bds x, t) \in 
 \mbb T^2 \times (0, +\infty), \\
 \omega(\bds x, 0)  = \omega_0(\bds x)&, &\qquad &\bds x \in \mbb T^2, \label{eq:vort:eulerC} 
\end{align}
\end{subequations}
{in which} ${\omega_0 := \nabla^\perp \cdot \bds u_0}$ and
it will be assumed that $\int_{\TT^2} \omega_0(\x) \, d\x = 0$, such
that $\int_{\TT^2} \omega(t,\x) \, d\x = 0$ for all $t> 0$. Hereafter
our focus will be on the vorticity formulation
\eqref{eq:vort:euler}. We will refer to Sobolev spaces
{$H^m(\TT^2)$, $m \in \RR$,} with the inner product defined as
${\la f, g \ra_{H^m} := \int_{\mbb T^2} (1- \Delta)^m \bar{f} g
  \; d\bds x}$, where $\bar{\cdot}$ denotes complex conjugation, such
that the norm is given by
{$\| f \|_{H^m} = \sqrt{ \la f, g \ra_{H^m}}$} \citep{af05}.
Without loss of generality, we will focus our discussion on a subspace
of $H^m(\mbb T^2)$ consisting of zero-mean functions
\begin{equation}
{H^m_0(\mbb T^2) := \left\{f\in H^m(\mbb T^2), \int_{\mbb T^2}f \; d \bds x = 0 \right\}.}
\end{equation}
We will also use the space $L^2(\TT^2) := H^0(\TT^2)$. {In
  addition, we will consider Lebesgue and Sobolev non-inner-product
  spaces $L^p(\TT^2)$ and $W^{1,p}(\TT^2)$ with the norms
  $\| f \|_{L^p} := \left( \int_{\mbb T^2} | f |^p \, d\x
  \right)^{1/p}$ and
  $\| f \|_{W^{1,p}} := \| f \|_{L^p} + \| \bnabla f \|_{L^p}$ with
  $1 \le p < \infty$.}

Analysis of the stability of equilibrium solutions $\omega_s =
\omega_s(\x)$ of system \eqref{eq:vort:euler} is a classical subject
in mathematical fluid mechanics with general results {describing
  conditions under which flows become unstable.}
The metric chosen to measure
the deviation from the equilibrium captures different scales of
instability --- higher regularity spaces $H^m(\TT^2)$ with $m > 0$
register finer structures, such as filamentation, while the energy
space $H^{-1}(\TT^2)$ captures large-scale instabilities.
Koch's theorem \citep{Koch2002} states that in the finer sense, i.e.,
if the evolution of the vorticity $\omega(t)$ is measured in the
H\"older space $C^{k,\alpha}$, for any $k\in \mathbb{N}$ and $\alpha >
0$, any non-isochronous equilibrium in 2D is nonlinearly Lyapunov
unstable. {Here} ``non-isochronous'' means that all Lagrangian
trajectories in the equilibrium flow do not have the same period (a
typical example of an isochronous flow is solid-body rotation).

Most large scale instabilities are classically attributed to laminar
oscillatory structures as was established by the pioneering
Rayleigh-Fjortoft-Tollmien inflection point theory and its
contemporary operator theoretical formulations
\citep{Chandrasekhar1961,Drazin1981,Friedlander2000,Schmid2001book,Lin2005}.
In this case instability arises from a smooth unstable mode of the
linearized equation, which in turn gives rise to {nonlinear}
instability in the energy space by the full analogue of the Lyapunov
Theorem, see \citet{Friedlander1997,Lin2004} and references therein.
However, one of the simplest equilibrium solutions of
  \eqref{eq:euler}--\eqref{eq:vort:euler} to which this theory {\em
  does not} apply is the 2D Taylor-Green vortex which is
  defined as
\begin{equation}
\label{eq:TG}
\bds u_s(\x) = \left(-\cos(x_1)\sin(x_2), \sin(x_1)\cos(x_2)\right)^T, \qquad 
\omega_s(\x) = 2\cos(x_1)\cos(x_2)
\end{equation}
and features a doubly-periodic array of cellular vortices. Some
nontrivial generalizations of this equilibrium were recently
considered by \citet{ZhigunovGrigoriev2023}.

On the other hand, short-wavelength instabilities have been studied
using an asymptotic Wentzel-Kramers-Brillouin (WKB) approach borrowed
from geometric optics in which the solution of {the linearization
  of} \eqref{eq:euler} is represented as $\u(t,\x) = \a(t,\x,\bxi_0)
\exp\left[ i S(t,\x,\bxi_0) / \delta \right] + \O(\delta)$ for some
$\delta > 0$, where $\bxi := \nabla S$ is the wavenumber of the
perturbation and an analogous representation is used for the pressure
$p(t,\x)$ \citep{FriedlanderVishik1991,lh91}. Considering the
leading-order expressions obtained by plugging these ans\"atze into
{the linearization of} \eqref{eq:euler} and then taking the
asymptotic limit $\delta \rightarrow 0$ followed by switching to the
Lagrangian representation, one obtains a system of
ordinary-differential equations (ODEs) describing the evolution of the
Lagrangian coordinate $\x(t;\x_0)$, the perturbation wavenumber
$\bxi(t;\x_0,\bxi_0)$ and the amplitude of the perturbation
$\a(t;\x_0,\bxi_0,\a_0)$ in function of the corresponding initial
conditions $\x_0$, $\bxi_0$ and $\a_0$ (chosen such that $\bxi_0 \cdot
\a_0 = 0$ to ensure incompressibility). This system of ODEs, referred
to as the bicharacteristic problem, describes the time evolution of
oscillatory perturbations in the {short-wavelength} limit. An
instability of an equilibrium can then be detected if one can find a
solution of this system such that $|\a(t;\x_0,\bxi_0,\a_0)|$ grows in
time. While this approach makes it possible to conclude about an
instability of the equilibrium, given its local Lagrangian nature, it
does not provide any information about the global structure of the
instability in space.

The 2D Taylor-Green vortex \eqref{eq:TG} is one of {a number of
  exact solutions of the Euler equations known in a closed form. In
  the presence of viscosity $\nu$, the velocity field \eqref{eq:TG}
  gives rise to a closed-form solution of the Navier-Stokes system
  which decays in time as $\O\left(e^{-\nu t}\right)$.  Therefore, the
  Taylor-Green vortex often serves as a benchmark in computational
  fluid dynamics.}  Most of the investigations of the stability of the
2D Taylor-Green vortex have been carried out in the viscous setting
where \eqref{eq:TG} is not an exact equilibrium solution of the
Navier-Stokes system.  However, the main underlying assumption in
these studies was that the time scale on which the instabilities
develop is much shorter than $\O\left(e^{-\nu t}\right)$, the rate of
the viscous decay of \eqref{eq:TG}. These investigations typically
involved WKB analysis, solution of the eigenvalue problem for the
linearized operator and/or time-integration of the governing
equations, all of which were performed numerically.  They included
analysis of elliptic instabilities under 3D perturbations
\citep{SippJacquin1998,Kerswell2002, AspdenVanneste2009} and
hyperbolic instabilities
\citep{LeblancGodeferd1999,FriedlanderVishik1991,Suzuki2018}.  A
thorough discussion of different instability mechanisms possible in
the Taylor-Green vortex under rotation and/or stratification can be
found in \cite{HattoriHirota2023}.  The elliptic instability is
closely related to elliptic stagnation points in the base flow and
only occurs when the perturbation is three-dimensional, crucially
depending on the wavenumber of the perturbation along the direction
orthogonal to the plane of motion.  On the other hand, the hyperbolic
instability is connected to hyperbolic stagnation points and appears
even under two-dimensional perturbations \citep{GauHattori2014}, which
is more relevant to the current study where we focus on analyzing the
stability of 2D flows. In particular, in this latter study the authors
considered a problem similar to the one investigated here.  However,
as will be evident below, our findings are in fact quite different,
underlining the difference between the viscous and inviscid
formulations.

In contrast to these earlier studies, our focus here is on the
instability of the Taylor-Green vortex \eqref{eq:TG} in 2D inviscid
Euler flows governed by \eqref{eq:euler}--\eqref{eq:vort:euler}.  Even
though at a formal level the Euler equations {under periodic
  boundary conditions} can be viewed as the vanishing viscosity limit
of the Navier-Stokes equations, the spectra of the corresponding
linearized operators are fundamentally different.  Unlike the
linearized Navier-Stokes operator, the linearized Euler operator is
not elliptic, thus the existent theory about elliptic operators cannot
be applied. Moreover, it is also degenerate and non-self-adjoint,
which further complicates the analysis.  Most importantly, the
spectrum of the linearized Navier-Stokes operator defined on a bounded
domain subject to Dirichlet boundary conditions or periodic boundary
conditions can only consist of the discrete spectrum {and the
  corresponding eigenfunctions are smooth}.
\cite{ShvydkoyFriedlander2008} proved that the eigenvalues of the
linearized Navier-Stokes operator converge to unstable eigenvalues of
the linearized Euler operator which are outside the essential spectrum
as viscosity goes to zero, if such eigenvalues exist. However, despite
the simple structure of the 2D Taylor-Green vortex \eqref{eq:TG}, the
existence of unstable eigenvalues of the corresponding linearized
Euler operator is still an open question.  If such unstable
eigenvalues exist, {the regularity of the corresponding
  eigenfunctions is not a priori known and may be determined by the
  location of these eigenvalues relative to the essential spectrum
  \citep{Lin2005}.  } {Thus, due to these nuances, the inviscid
  problem is distinct from its viscous counterpart.}

Since the spectra of the linearized Euler operators obtained by
linearizing the velocity formulation \eqref{eq:euler} and the
vorticity formulation \eqref{eq:vort:euler} are equivalent
\citep{ShvydkoyLatushkin2005}, in this study, we use the latter
formulation and provide numerical evidence that the linearized
operator has unstable eigenvalues {approximately equal to
  $0.1424 \pm 0.5875 i$} with the corresponding {eigenfunctions}
given by {distributions} in {$H^{0.28}_0(\TT^2)$, where the
  level of regularity $s = 0.28$ is determined approximately based on
  Lin's theorem \citep{Lin2004},} {as will be discussed in
  \S\,\ref{sec:linear}.}  This eigenfunction exhibits a more regular
profile in the laminar cells loosing its smoothness in the vicinity of
the heteroclinic orbits of the equilibrium \eqref{eq:TG}. We also
illustrate another distinct instability mechanism associated with a
continuous family of uncorrelated functions corresponding to points in
the essential spectrum, which is quite different from the modal growth
observed in the former case.  Since the essential spectrum does not
arise in a finite-dimensional setting, investigation of these
questions requires the use of computational tools which are more
refined as compared to the techniques typically employed in the
studies of hydrodynamic stability \citep{Schmid2001book}. Obtaining
these results is enabled by the solution of a suitably defined PDE
optimization problem. Such optimization-based formulations have had a
long history in the study of flow stability problems, both linear and
nonlinear \citep{Schmid2001book,Kerswell2014,Kerswell2018}. However,
given the subtle infinite-dimensional nature of the optimization
problem considered here, we solve it using a specialized variant of
the adjoint-based approach which allows us to impose different levels
of regularity on the obtained optimal initial conditions
\citep{ZhaoProtas2023}. By solving this optimization problem using
increasing spatial resolutions, we obtain a sequence of functions that
are localized near the hyperbolic stagnation points of the equilibrium
solution \eqref{eq:TG} and reveal high-frequency oscillations
restricted by the spatial resolution.  Importantly, using these
functions as initial conditions, the corresponding solutions of the
linearized Euler system reveal growth rates saturating rigorous a
priori bounds on the growth of the semigroup induced by the essential
spectrum of the generator.  While these results are consistent with
the findings of the WKB analysis, they also provide information about
the global spatial structure of the perturbations realizing this
maximum possible growth.

  The structure of the paper is as follows: in \S\,\ref{sec:linear} we
  introduce the {problems of linear and nonlinear} instability,
  and discuss the spectrum of the linearized Euler operator; in
  \S\,\ref{sec:numer} we discuss the numerical discretization of the
  linearized operator to compute its eigenvalues as well as the
  formulation of a PDE optimization problem to obtain initial
  conditions such that the corresponding flows realize the largest
  growth rate of perturbations predicted by the form of the essential
  spectrum, which is solved using a Riemannian conjugate gradient
  method described in Appendix \ref{sec:RCG}; in
  \S\,\ref{sec:results}, we illustrate two distinct mechanisms that
  lead to a linear instability --- a modal growth and a nonmodal
  growth of the solution, where the former corresponds to the point
  spectrum while the latter corresponds to the essential spectrum of
  the linearized operator and is highly dependent on the
  {function} space in which the perturbation is defined; in that
  section we also discuss some computational results concerning the
  nonlinear instability; discussion and final conclusions are deferred
  to \S\,\ref{sec:final}.

\section{Linear {and nonlinear} stability}
\label{sec:linear}

Linearizing system \eqref{eq:vort:euler} around a steady solution
$\{\bds u_s, \omega_s\}$, we obtain the following system
\begin{subequations} 
\label{eq:vort:linear}
\begin{align}
\partial_t w = \L w &, &\qquad &(\bds x, t) \in 
 \mbb T^2 \times (0, +\infty), \\
 w(\bds x, 0)  = w_0(\bds x)&, &\qquad &\bds x \in \mbb T^2, \label{eq:vort:linear0}
\end{align} 
\end{subequations}
where the linearized Euler operator $\L$ is given by
\begin{equation}
\L w := -(\bds u_s\cdot \nabla) w -({\bds u} \cdot \nabla)\omega_s
= {\left[ -\bds u_s \cdot \nabla - \nabla\omega_s\cdot (\nabla^\perp \Delta^{-1}) \right] w}.
\label{eq:L}
\end{equation} 
Solution of system \eqref{eq:vort:linear} can be written as
$w(t) = e^{t \L} w_0$, where $w(t) := w(t,\cdot)$ and $e^{t \L}$ is
the semigroup induced by the operator $\L$ \citep{EngelNagel2000}. The
question of stability of the equilibrium $\omega_s$ is thus linked to
the asymptotic, as $t \rightarrow \infty$, behavior of
$\| e^{t \L} \|_{H^m}$ quantified by the growth abscissa
$\gamma(\L) := \lim_{t \rightarrow \infty} t^{-1} \ln \| e^{t \L}
\|_{H^m}$, which is in turn determined by the spectrum $\sigma(\L)$ of
the operator $\L$. While in finite dimensions it is determined by the
eigenvalue with the largest real part, in infinite dimensions the
situation is more nuanced since there exist operators $\A$ such that
$\sup_{z \in \sigma(\A)} \Re(z) < \gamma(\A)$, e.g., {Zabczyk's problem
\citep{Zabczyk1975, Trefethen1997}}; some
problems in hydrodynamic stability where such behavior was identified
are analyzed by \citet{Renardy1994}.

Following \citet{Browder1961}, we decompose the spectrum of
$\L$ into two disjoint sets: the discrete spectrum and the essential
spectrum, as follows,
\begin{equation}
\sigma(\L) = \sigma_{\disc}(\L) \ \cup \ \sigma_{\ess}(\L).
\end{equation}
We then say that $z \in \sigma_\disc(\L)$ if it satisfies the following conditions:
\begin{itemize}
    \item[(i)] \ $z$ is an isolated point in $\sigma(\L)$;
    \item[(ii)] \ $z$ has finite multiplicity, i.e., $\bigcup_{r=1}^\infty
    \mathrm{Ker}(z - \L)^r$ is finite dimensional;
    \item[(iii)] \ the range of $z-\L$ is closed.
\end{itemize}
Otherwise, $z$ is called a point of the essential spectrum $\sigess(\L)$.
{To illustrate this concept, we consider the linear operator $T$ that maps all functions in $L^2_0(\mbb T)$ to the zero
function. It has only the essential spectrum $\sigma_\ess(T) = \{0\}$ as
its kernel
$\mathrm{Ker}(T) = L^2_0(\mbb T)$ is infinite-dimensional.
As a more complicated example, we consider the linear operator $T: L^2_0(\mbb T) \to L^2_0(\mbb T)$ defined by
\begin{equation}
T[f](x) := \sum_{n=1}^\infty \frac{1}{n}\left( a_n\cos(nx) +
  b_n\sin(nx) \right), \quad \text{where} \quad f(x) =  \sum_{n=1}^\infty (a_n\cos(nx) + b_n\sin(nx))
\end{equation}
For any positive integer $p$, $1/p$ is an eigenvalue of $T$, and
$\cos(p x)$ and $\sin(p x)$ are the corresponding eigenfunctions.
Since $T$ is not surjective, $0 \in \sigma(T)$ and since
$\lim_{p\to\infty} (1/p) = 0$, 0 is not an isolated point in the
spectrum of $T$. Therefore, we have
$\sigma_{\disc}(T) = \left\{1/p, \;  p \in \mbb N_+\right\}$ and
$\sigma_{\ess}(T) = \{0\}$.}

While in finite dimensions linear operators can be represented as
matrices which can only have a discrete spectrum, in infinite
dimensions the situation is complicated by the presence of the
essential spectrum. We refer to the set of eigenvalues of $\L$ as the
point spectrum
\begin{equation}
{\sigp(\L) := \left\{ \lambda \in \CC \; : \; \exists \phi (\x) \neq 0,  \quad \L \phi (\x) = \lambda \phi (\x), \quad \x \in \TT^2  \right\}},
\label{eq:sig0}
\end{equation}
where $\phi$ is the eigenfunction corresponding to the eigenvalue
$\lambda$. 
{It follows from the discrete translation symmetry
  of the 2D Taylor-Green vortex \eqref{eq:TG} and the continuous
  translation invariance of the Euler system \eqref{eq:euler}, that if
  $\phi(x_1, x_2)$ is an eigenfunction corresponding to $\lambda$,
  then so is $\phi(-x_1, -x_2)$, whereas $\phi(x_1+\pi, x_2)$,
  $\phi(x_1, x_2+\pi)$, $\phi(-x_1, x_2)$, and $\phi(x_1, -x_2)$ are
  eigenfunctions corresponding to $-\lambda$.}

As regards the {discrete spectrum ${\sigma_\disc}(\L)$} of the linearized
Euler operator \eqref{eq:vort:linear}, some results are available only
for certain flows such as parallel and rotating shear flows
\citep{Drazin1981, Chandrasekhar1961, Friedlander1997} and the
cellular ``cat's-eye'' flow \citep{Friedlander2000}.  
In the absence of general results, one of the goals of the present study is to
consider this question in the context of the Taylor-Green vortex
\eqref{eq:TG}. Unlike the aforementioned two cases where the
instability is closely related to the shear flow structure of the
equilibria, equilibrium \eqref{eq:TG} possesses a cellular
structure only.


On the other hand, the essential spectrum $\sigess(\L)$ of the linear
operator $\L$ is fully understood \citep{ShvidkoyLatushkin2003}: in
$H^{m}_0(\TT^2)$, {$m \in \RR$,} it is given by the strip
\begin{equation}
  \sigess(\L; H^{m}_0) = \{ z\in \CC: |{\Re}(z)| \leq |m| \, \mumax \},
  \label{eq:sigess}
\end{equation}
where $\mumax$ is the maximal Lyapunov exponent corresponding
to the Lagrangian flow $\varphi_t: \bxi \to \bds x(t; \bxi)$ generated
by the steady state via $\partial_t \bds x(t) = \bds u_s (\bds x(t))$,
\begin{equation}
{\mumax} =  \lim_{t\to\infty} \frac{1}{t}\log \sup_{\bds x\in\mbb T^2} ||\nabla \varphi_t(\bds x)||. 
\end{equation}
In 2D, $\mumax$ can only be attained at a hyperbolic stagnation
point $\x_s$ of the flow $\{\varphi_t\}$ induced by the steady state
$\u_s$ and is determined by the largest real part of the {eigenvalues}
of the velocity gradient $\bnabla\u_s({\x_s})$ evaluated over all
stagnation points ${\x_s}$
\citep{ShvydkoyFriedlander2005}. {The} equilibrium state
\eqref{eq:TG} has four hyperbolic stagnation points
$\x_s = \left\{(\pi/2, \pi/2), (\pi/2, 3\pi/2), (3\pi/2, \pi/2),
  (3\pi/2, 3\pi/2)\right\}$. By computing the eigenvalues of
$\bnabla \u_s$ at these four points, we deduce that ${\mumax} =
1$. {Another interesting property of the stagnation points is
  that the action of the linearized operator on any sufficiently
  smooth function $w$ vanishes at these points, i.e.,
  \begin{equation}
    \L w = 0 \qquad \text{at} \ \x = \x_s.
    \label{eq:Lwxs}
  \end{equation}}
  

At the same time, we also have 
\begin{equation}
\sigma(e^{t\L}; H^{m}) = \{ z \in \CC:  e^{-t|m|} \leq |z| \leq e^{t|m|} \},
\label{eq:SMT}
\end{equation}
such that the full analogue of the Spectral Mapping Theorem holds
\citep{ShvidkoyLatushkin2003}.  All points in the band and the
annulus, respectively, are points of the essential spectrum in the
Browder sense \citep{Browder1961}, which is the broadest
  definition of the essential spectrum also coinciding with the
Fredholm spectrum.  In the proof, for any point $z\in\sigess(\L)$,
\cite{ShvidkoyLatushkin2003} constructed approximate
  eigenfunctions as a sequence of unit vectors $\{f_n\} \in
    H^m_0(\TT^2)$ such that $\|(\L - z) f_n \|_{H^m} \rightarrow 0$
as $n \rightarrow \infty$, and $\{f_n\}$ does not contain any
convergent subsequence.  These approximate eigenfunctions are
  characterized by highly oscillatory behavior and are stretched
along the heteroclinic orbits of $\bds u_s$ while
concentrating towards the hyperbolic points. These results are
  consistent with the asymptotic WKB analysis conducted in the
neighbourhood of the hyperbolic stagnation points which suggests the
presence of highly oscillatory perturbations growing as $\O(e^{\mumax
  t})$, though they need not be eigenfunctions of $\L$
\citep{FriedlanderVishik1991,lh91}. In general, it is unknown
whether the operator $\L$ has any unstable
eigenvalues. {However, when it does, the regularity of the
  corresponding eigenfunctions is characterized by a theorem of
  \citet{Lin2004} which we state here in a slightly less general
  version adapted to the case when the equilibrium is given by the
  Taylor-Green vortex \eqref{eq:TG}.}
  {
  \begin{thm}[\citet{Lin2004}]
    \label{thm:Lin}
     Suppose there exists an exponentially growing solution
  $e^{\lambda t} w_0$ of the linearized system $\partial_t w = \L w$
  with $\Re(\lambda) > 0$ and let $w_0 \in L^2(\TT^2)$. Then we have
  the following
\begin{itemize}
\item[(i)] [regularity of growing modes] $w_0 \in W^{1,p}(\TT^2)
  \cap L^q(\TT^2)$ for all $1 \le p < p^*$ and $1 \le q < \infty$,
  where
  \begin{equation}
    p^* = \begin{cases}
      \frac{\mumax}{\mumax - \Re(\lambda)} = \frac{1}{1 -\Re(\lambda)}, \qquad & \mumax > \Re(\lambda), \\
      \infty, & \mumax \le \Re(\lambda),
    \end{cases}
    \label{eq:p*}
    \end{equation}

\item[(ii)] [nonlinear instability] For any $p \in [1,p^*)$, $q\in[1,\infty)$, $m\in[-1,\infty)$, there exists $\epsilon > 0$,
    such that  for any $\delta > 0$, there is a solution $\omega^\delta(t)$
    of the 2D Euler system \eqref{eq:vort:euler} corresponding to the initial condition
    $\omega_0^{\delta}$, satisfying
    \begin{equation*}
\left\| \omega_0^{\delta} - \omega_s \right\|_{L^q} +  \left\| \bnabla\left(\omega_0^{\delta} - \omega_s\right) \right\|_{L^p} \le \delta,
\end{equation*}
and
\begin{equation*}
\sup_{0 < t < T_\delta} \left\|\omega^\delta(t) - \omega_s\right\|_{H^m} \geq \epsilon.
\end{equation*}
\end{itemize}
\end{thm}}
\noindent
{While for general infinite-dimensional nonlinear systems linear
instability need not imply a nonlinear instability, the second part of
the theorem above asserts that this is in fact the case for the 2D
Euler problem, provided the unstable eigenfunction of the linearized
operator $\L$ is sufficiently regular.  }

{As a key result of the present study, we provide numerical
  evidence that the operator $\L$ does possess unstable eigenvalues
  and we also characterize the regularity of the corresponding
  eigenfunctions} {concluding that it is consistent with Theorem
  \ref{thm:Lin}, part (i), cf.~\S\, \ref{sec:modal_growth}.} {The
  nonlinear instability predicted in part (ii) of the theorem is
  illustrated in \S\,\ref{sec:nonlinear_growth}.}  Another
contribution of the present study is to illustrate the nontrivial
instability mechanism associated with the unstable essential spectrum,
cf.~\S\, \ref{sec:nonmodal_growth}.

\section{Numerical approaches}
\label{sec:numer}

In this section we introduce the numerical approaches that will allow
us to characterize the growth of solutions of the linear and nonlinear
problems \eqref{eq:vort:linear} and \eqref{eq:vort:euler}. First, in
\S\,\ref{sec:point}, we describe a numerical solution of {the} eigenvalue
problem \eqref{eq:sig0} such that the eigenfunctions $\phi$
corresponding to the eigenvalues $\lambda \in \sigma_{p}(\L)$ can be
used as the initial condition in the linear and nonlinear problems
\eqref{eq:vort:linear} and \eqref{eq:vort:euler} (in the latter case
the eigenfunctions serve as perturbations of the equilibrium
\eqref{eq:TG}). Then, in \S\,\ref{sec:ess}, we introduce an
optimization-based approach allowing us to construct solutions of the
linear problem \eqref{eq:vort:linear} saturating the spectral bounds
\eqref{eq:sigess} and \eqref{eq:SMT}. Finally, in \S\,\ref{sec:PDE},
we describe the approach to the numerical solution of the evolutionary
systems \eqref{eq:vort:euler}, \eqref{eq:vort:linear} and its adjoint.

\subsection{The point spectrum of the linear operator $\L$}
\label{sec:point}

To characterize the point spectrum ${\sigp(\L)}$, {we} adopt
a Galerkin approach where the operator $\L$ is discretized using the
following orthonormal basis in ${H^m_0(\TT^2)}$
\begin{subequations}
\label{eq:varphipsi}
\begin{align}
\varphi_{j_1, j_2}(\x)  & := \phantom{-}\frac{1}{\sqrt{2}\pi} (1 + j_1^2 + j_2^2)^{-m/2}\cos(j_1 x_1 + j_2 x_2), \qquad j_1, j_2 \in \mbb N,\label{eq:varphi} \\
\psi_{j_1, j_2}(\x) & := -\frac{1}{\sqrt{2}\pi} (1 + j_1^2 + j_2^2)^{-m/2} \sin(j_1 x_1 + j_2 x_2), \label{eq:psi} 
\end{align}
\end{subequations}
and we have
\begin{equation}\label{eq:L:action}
\begin{gathered}
\L \varphi_{j_1,j_2} =  \alpha (\varphi_{j_1+1, j_2+1}  -  \varphi_{j_1-1, j_2-1})
+ \beta (-\varphi_{j_1+1, j_2-1} +  \varphi_{j_1-1, j_2+1}), \\
\L \psi_{j_1,j_2} =  \alpha (\psi_{j_1+1, j_2+1}  -  \psi_{j_1-1, j_2-1})
+ \beta (-\psi_{j_1+1, j_2-1} +  \psi_{j_1-1, j_2+1}),\\
\text{where} \qquad
\alpha = \frac{(j_1-j_2)(j_1^2+j_2^2-2)}{4(j_1^2+j_2^2)}, \qquad 
\beta = \frac{(j_1+j_2)(j_1^2+j_2^2-2)}{4(j_1^2+j_2^2)}.
\end{gathered}
\end{equation}
In the computations, we approximate functions in $H^m_0(\TT^2)$ using
a finite subset of the basis \eqref{eq:varphipsi}
\begin{equation}
\W^N = \left\{\varphi_{0, j_2}, \psi_{0, j_2}: 1\leq j_2 \leq N\right\} 
\bigcup \left\{\varphi_{j_1, j_2}, \psi_{j_1, j_2}: 1\leq j_1 \leq N, -N\leq j_2\leq N\right\},
\end{equation}
which contains $|\W^N| = 2\sum_{s = 1}^N 4s = 4N(N+1)$ elements. We
label the basis functions in $\W^N$ using the ``spiral'' ordering,
i.e.,
\begin{equation}
\begin{aligned}
p_{n+2j} &= \varphi_{j, s},  &p_{n+2j+1} &= \psi_{j, s}, \\
p_{n+2s+2j} &= \varphi_{s, s-j},  &p_{n+2s+2j+1} &= \psi_{s, s-j},\\
p_{n+4s+2j} &= \varphi_{s, -j},  &p_{n+4s+2j+1} &= \psi_{s, -j}, \\
p_{n+6s+2j} &= \varphi_{s-j, 1-s},  &p_{n+6s+2j+1} &= \psi_{s-j, 1-s},
\end{aligned}
\qquad 
\begin{gathered}
n = 1+4s(s-1), \\ 0\leq j\leq s-1, \; 1\leq s \leq N.
\end{gathered}
\end{equation}
Given a function  $f \in H_0^m(\TT^2)$, we thus define its Galerkin approximation $f^N$ by
\begin{equation}
  f \approx f^N := \sum_{j = 1}^{|\W^N|} \widehat{f}_j p_j,
  \label{eq:fhat}
 \qquad 
 \widehat{f}_j = \la f, p_j \ra_{H^m}.
 \end{equation}
 Approximating the eigenfunctions $\phi$ in \eqref{eq:sig0} in terms of
 the truncated Fourier series \eqref{eq:fhat}, we arrive at the
 discrete algebraic eigenvalue problem
\begin{equation}
\bL \, \phi = \lambda \phi,
\label{eq:evpN}
\end{equation}
where $\bL$ is a ${|\W^N|\times|\W^N|}$ matrix whose entries are determined by
relations \eqref{eq:L:action} as
\begin{equation}
\bL_{j,k} = \left\langle \L p_k, \; p_j \right\rangle_{H^m}, \qquad 1 \leq j, k \leq {|\W^N|}.
\end{equation}
{As a result of relations \eqref{eq:L:action}, matrix $\bL$ is
  sparse, with at most four nonzero entries in each row and column. }
Moreover, since
\begin{equation}
\begin{aligned}
\lla \L p_k, \; p_j\rra_{H^m} &= 
{\lla (1-\Delta^{-1})^{m/2}\L p_k, \;(1-\Delta^{-1})^{m/2}p_j\rra_{L^2}}\\
&= {\lla \left[(1-\Delta^{-1})^{m/2}\L(1-\Delta^{-1})^{-m/2}\right] (1-\Delta^{-1})^{m/2} p_k, \;
(1-\Delta^{-1})^{m/2} p_j\rra_{L^2},}
\end{aligned}
\end{equation}
the matrices $\bL$ computed in different Sobolev spaces $H_0^m(\TT^2)$
are similar. Therefore, without loss of generality, we can focus our
discussion on the matrix constructed with $m = 0$, i.e., in
$L^2_0(\TT^2)$.  

{We adopt two different methods to {numerically solve the
    algebraic eigenvalue problem \eqref{eq:evpN}. As} the first
  method, we use the eigenvalue solver \texttt{dgeev} from the LAPACK
  library to compute {\em all} eigenvalues of $\bL$. This approach
  provides a complete picture of the spectrum of the matrix
    $\bL$, but is computationally expensive, limiting the resolution
  to $N^2 = 200^2$. The second method takes advantage of the sparse
  structure of the matrix $\bL$ and uses a Krylov subspace
  method \citep{HattoriHirota2023} to only compute the eigenvalue with
  the largest real part and the corresponding
    eigenvector. Specifically, we use the Matlab function
  \texttt{eigs}, setting the dimension of the Krylov subspace to 20,
  the tolerance to $10^{-10}$, and the maximum number of
    iterations to 1000.  To validate these results, at $N^2 = 200^2$,
    we use a random vector to generate the Krylov subspace, and the
    obtained eigenvalues with the largest real part are found to be
    essentially the same as the ones obtained using the LAPACK
    subroutine \texttt{dgeev}.  To speed up the computation, at the
    resolution $(2N)^2$, we use $\lambda_+^N$ as the shift and the
    corresponding eigenfunction $\phi_+^N$ as the generator of the
    Krylov subspace.  This allows us to increase the numerical
  resolution from $N^2 = 200^2$ to $3000^2$.} {A combination of
  these two approaches makes it possible to obtain a global picture of
  the spectrum of the matrix $\bL$ while also refining the
  approximations of the most interesting eigenvalues.}

As is shown in \S\,\ref{sec:linear_growth}, employing the procedure
described above, we obtain unstable eigenvalues whose real part is
around {0.1424} and the corresponding eigenfunctions belong to
{$H^{0.28}_0(\TT^2) \subset L^2_0(\TT^2)$}. Using the real part
of this eigenfunction as the initial condition in the linearized Euler
equations \eqref{eq:vort:linear}, we observe an exponential growth of
the {$L^2$} norm of the solution $w(t)$ with the rate predicted
by the real part of the unstable eigenvalue.  However, as is evident
from \eqref{eq:sigess}, in {the Sobolev spaces $H^1$ and
  $H^{-1}$}, $\sigess(\L)$ forms a vertical band
$|{\Re}(z)| \le 1$. It is thus a natural question what initial
condition can realize the growth abscissa $\gamma(\L) = 1$ predicted
by $\sigess(\L)$, which is larger than the growth rate realized by the
unstable eigenfunction. Tools needed to address this question are
discussed next.

\subsection{The essential spectrum of the linear operator $\L$}
\label{sec:ess}


As will be evident in \S\,\ref{sec:linear_growth}, {the real part
  of the unstable eigenvalues of problem \eqref{eq:sig0} found as
  described in \S\,\ref{sec:point} is near 0.1424,} and therefore, for
$m \neq 0$, does not saturate the bounds on the growth of the abscissa
implied by \eqref{eq:SMT}. It is therefore natural to ask the question
whether there exists an initial condition $w_0$ such that the growth
rate of $||e^{t\L} w_0||_{H^m}$, i.e.,
$(d/dt) \ln(\| e^{t\L} w_0 \|_{H^{m}})$, saturates this bound.  Since
the essential spectrum is an inherently infinite-dimensional object,
information about it is lost in a finite-dimensional truncation such
as \eqref{eq:fhat}. We thus need an approach different from the method
described in \S\,\ref{sec:point} to study properties related to the
essential spectrum. Instead of maximizing the growth rate of the
solutions of \eqref{eq:vort:linear} directly, we aim to maximize the
norm of the solution $||e^{t\L} w_0||_{H^m}$ at some finite time
$t = T > 0$ over all $w_0 \in H^m_0(\TT^2)$.  Since $e^{t \L} w_0$ is
linear with respect to $w_0$, we can fix $\|w_0\|_{H^m} = 1$ without
loss of generality.  Therefore, we define the following objective
functional $J \; : \; {H_0^m(\TT^2)} \rightarrow \RR$
\begin{equation}
  J (w_0) = \left|\left|e^{T\L} w_0 \right|\right|^2_{H^m},
  \label{eq:J0}
\end{equation}
and the corresponding optimization problem:
\begin{problem}\label{prob:ess:growth}
For $T > 0$, find
\begin{equation}
{\wopt_0 = \argmax_{w_0 \in \mc M} 
J (w_0),  \qquad \mc M := \{w_0 \in H^m_0(\TT^2):  \|w_0\|_{H^m} = 1\}.}
\end{equation}
\end{problem}
\noindent
To observe a significant exponential growth of ${||e^{T\L}
w_0||_{H^m}}$, one normally chooses $T > O(\ln (\|w_0\|_{H^m}))$,
and in our study we use $T = 1$. {Problem \ref{prob:ess:growth}
  has the form of a quadratically-constrained quadratic program
  defined in terms of positive-semidefinite operators and is therefore
  convex.}

While discretized versions of Problem \ref{prob:ess:growth} can
  in principle be solved by performing a singular-value decomposition
  of the corresponding matrix exponential \citep{Schmid2001book}, this
  is problematic when one has to ensure the required regularity of the
  optimal initial condition $\wopt_0$, which is encoded here
  in the choice of $m$. We therefore solve this problem using a
Riemannian conjugate gradient method \citep{ams08,DanailaProtas2017,
  Sato:2021:RCG,ZhaoProtas2023} which requires the computation the
Sobolev gradient of ${J(w_0)}$, denoted ${\nabla J(w_0)}$. Evaluating
the G\^{a}teaux (directional) differential ${J'(w_0; w_0'): H_0^m
  \times H^m_0 \to \mbb R}$, which represents the variation of the
objective function ${J(w_0)}$ in the direction of ${w'_0}$ at the
point ${w_0}$, we obtain
\begin{equation}
\begin{aligned}
J'(w_0; w_0') &= \lim_{\epsilon \to 0} \frac{1}{\epsilon} \left[J(w_0 + \epsilon w_0') - J(w_0) \right]\\
& = 2 \left\la e^{T\L} w_0, \; e^{T\L} w_0' \right \ra_{H^m} \\
& = {2 \left\la (1-\Delta^{-1})^m e^{T\L} w_0, \; e^{T\L} w_0' \right \ra_{L^2}} \\
& = {2 \left \la  e^{T\L^*} (1-\Delta^{-1})^m e^{T\L} w_0, \; w_0' \right \ra_{L^2}} \\
& = {2 \left \la  (1-\Delta^{-1})^{-m} e^{T\L^*} (1-\Delta^{-1})^m e^{T\L} w_0, \; w_0' \right \ra_{H^m},}
\end{aligned}
\label{eq:dJ}
\end{equation}
where $\L^*$ is the adjoint of the linear operator $\L$ defined with
respect to the $L^2$ inner product as $\langle\L f, g \rangle_{L^2}
=\langle f, \L^*g \rangle_{L^2}$ and having the form
\begin{equation}
  {\L^* = \u_s\cdot \nabla + \Delta^{-1}(\nabla w_s \cdot \nabla^\perp).}
  \label{eq:L^*}
\end{equation}
Finally, using relation \eqref{eq:dJ} and the Riesz
representation theorem, the Sobolev gradient of ${J(w_0)}$ with respect to
the $H^m$ inner product is obtained as
\begin{equation}
{\nabla J(w_0) = 2 (1-\Delta^{-1})^{-m} e^{T\L^*} (1-\Delta^{-1})^m e^{T\L} w_0.}
\label{eq:gradJ2}
\end{equation}
Details of the Riemannian conjugate gradient method we use to
solve Problem \ref{prob:ess:growth} are described in Appendix
\ref{sec:RCG}.

\subsection{Numerical Solution of the Evolution Problems}
\label{sec:PDE}

Here we describe the numerical approach we use to solve the evolution
problems \eqref{eq:vort:euler}, \eqref{eq:vort:linear} and the adjoint
problem defined in \eqref{eq:L^*}. We employ a standard
Fourier-Galerkin pseudospectral method \citep{Canuto1993book} where
the solution is approximated in terms of a truncated Fourier series
with the nonlinear term and the terms with nonconstant coefficients
evaluated in the physical space. In lieu of dealiasing we use the
Gaussian filter proposed by \citet{hl07}. The resulting system of
ordinary differential equations is integrated in time using the RK4
technique and a massively parallel implementation based on {\tt MPI}.
Since the considered initial conditions are distributions, rather than
smooth functions, cf.~figure \ref{fig:eigfun}, the numerical solutions
of problems \eqref{eq:vort:euler} and \eqref{eq:vort:linear}
are not well resolved regardless of the resolution {$N^2$}. However, the
Galerkin projection implied by the truncation of the series as in
\eqref{eq:fhat} together with the resolution-dependent filter can be
regarded as a regularization of the problem whose effect vanishes as
the resolution is refined, i.e., as $N \rightarrow \infty$.

\section{Results}
\label{sec:results}

Here we describe the mechanisms of the linear growth of perturbations
in the {\em modal} regime, associated with eigenvalues in the point
spectrum $\sigp(\L)$, and in the {\em nonmodal} regime, associated
with points in the essential spectrum $\sigess(\L)$
that do not coincide with the point spectrum ${\sigp(\L)}$. This is followed
by a discussion of the growth of perturbations in the nonlinear
regime. Hereafter, we will use the convention that the superscript $N$
will represent the resolution with which a given quantity, such as an
eigenvalue, eigenfunction or a solution of the linear problem
\eqref{eq:vort:linear}, is approximated.

\subsection{Linear {instability}}
\label{sec:linear_growth}

\begin{figure}
  \centering
  \begin{subfigure}{0.48\textwidth}
    \includegraphics[width=\textwidth]{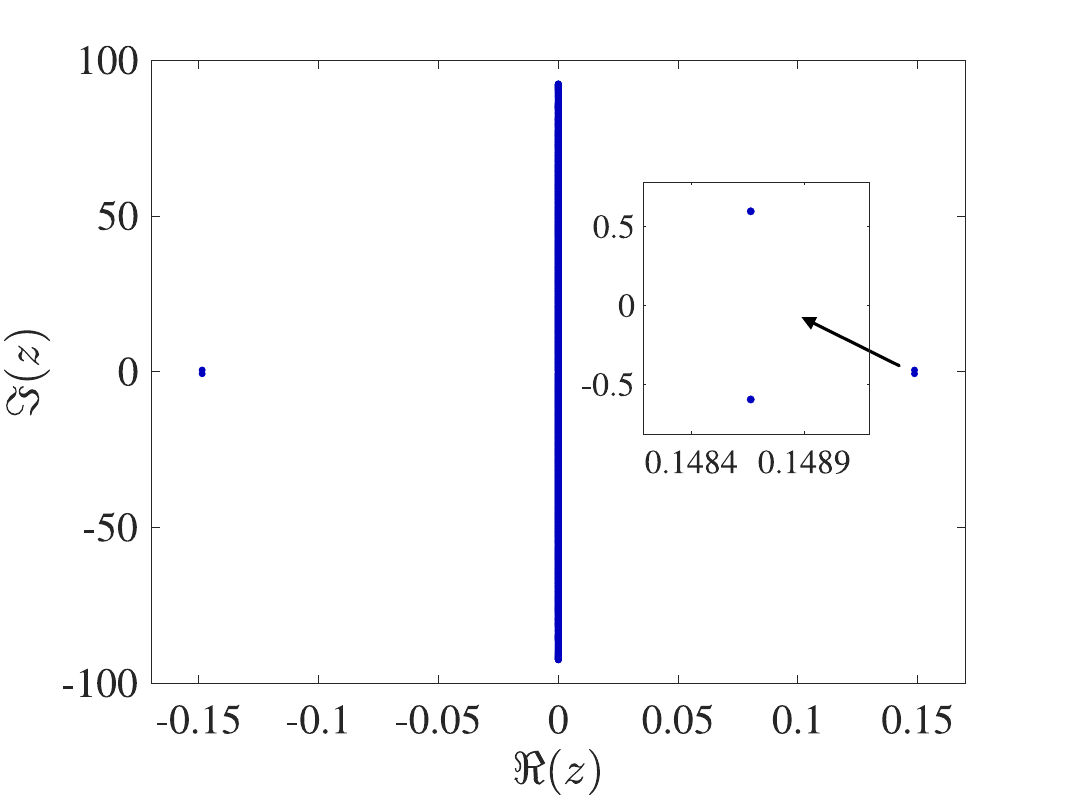}
    \caption{}
  \end{subfigure}
  \hfill
  \begin{subfigure}{0.48\textwidth}
    \includegraphics[width=\textwidth]{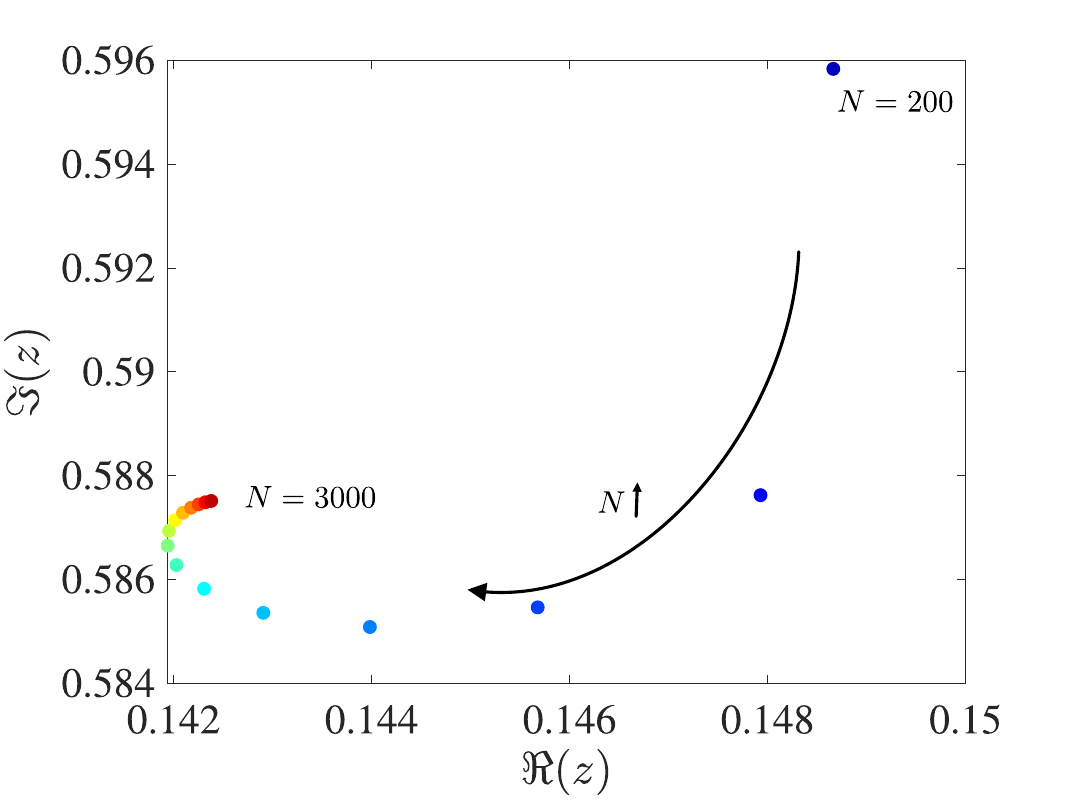}
    \caption{}
  \end{subfigure}

 \begin{subfigure}{0.48\textwidth}
   \includegraphics[width=\textwidth]{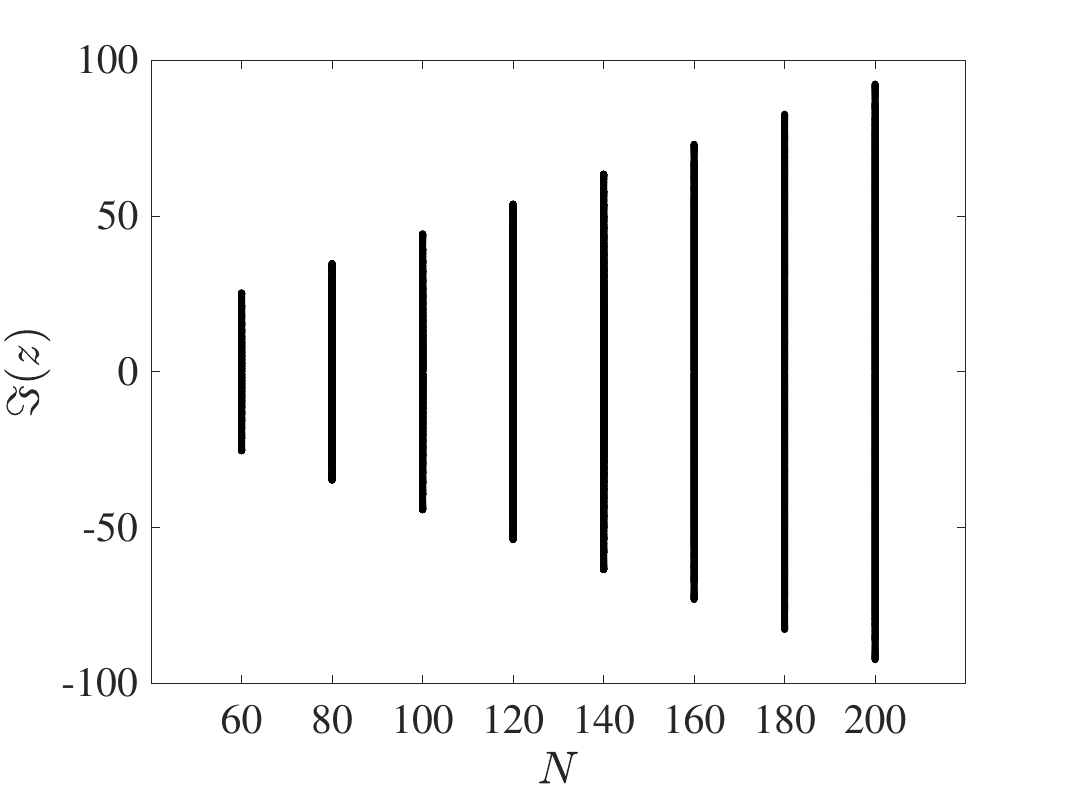}
    \caption{}
    \end{subfigure}
    \hfill
    \begin{subfigure}{0.48\textwidth}
      \includegraphics[width=\textwidth]{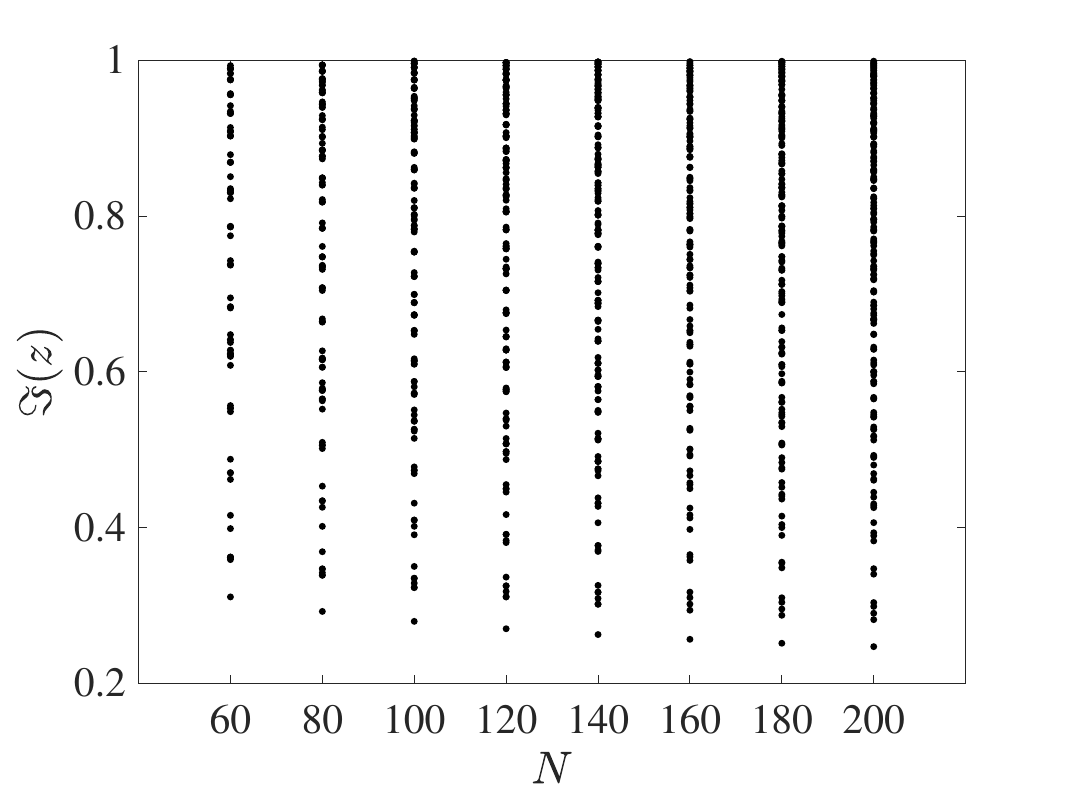}
      \caption{} 
  \end{subfigure}
  \caption{(a) {Eigenvalues of the discrete eigenvalue problem
      \eqref{eq:evpN} obtained with $m= 0$ and the resolution
      $N^2 = 200^2$ (the inset represents a magnification of
        the neighbourhood of the unstable eigenvalues $\lambda_+^{200}$
        and $\overline{\lambda_+^{200}}$). (b) The eigenvalue
      $\lambda_+^N$ computed using the Krylov subspace method with
      resolutions $N^2 = 200^2, 400^2, \ldots, 3000^2$.}  (c)
    Imaginary parts of the remaining eigenvalues
    $\Im\left(\lambda^N\right)$ obtained for different resolutions
    $N^2$ with panel (d) showing a magnification of a region near the
    origin.}
  \label{fig:evals}
\end{figure}

As discussed in \S\,\ref{sec:point}, since in the discrete eigenvalue
problem \eqref{eq:evpN} the matrices corresponding to different values
of $m$ are similar, it suffices to solve the eigenvalue problem using
$m = 0$ only.  {Figure \ref{fig:evals}a shows the eigenvalues of
  the discrete eigenvalue problem \eqref{eq:evpN} with $m = 0$
  obtained using the resolution $N^2 = 200^2$.} We see that there are
two pairs of conjugate eigenvalues {$\pm \lambda_+^{200},
  \overline{\pm \lambda_+^{200}}$, where $\lambda_+^{200}$} denotes
the eigenvalue whose real and imaginary parts are both positive.
{In order to better resolve these eigenvalues and the
  corresponding eigenvectors, the discrete eigenvalue problem
  \eqref{eq:evpN} is then solved with the Krylov method described in
  \S\,\ref{sec:point} which leverages the sparsity of the matrix
  $\bL$. This allows us to refine the resolution as $N^2 = 200^2,
  400^2, \ldots 3000^2$ and the obtained eigenvalues $\lambda_+^N$ are
  shown in figure \ref{fig:evals}b. We see that,} as the resolution
${N^2}$ increases, these eigenvalues converge to {a well-defined
  limit; this limit is interpreted as the ``true'' eigenvalue} in the
point spectrum ${\sigp(\L)}$ {\citep{Boyd2001}}.  We denote
$\lim_{N\to\infty} \lambda_+^N =: \lambda_+$ and at the largest
resolution {$N^2 = 3000^2$} have {$\lambda_+^{3000} = 0.1424
  + 0.5875i$} which is a numerical approximation of the ``true''
unstable eigenvalue $\lambda_+$.  {We note that 0 is also an
  eigenvalue.}  On the other hand, all remaining eigenvalues of the
{discretized} problem \eqref{eq:evpN} fall on the imaginary axis and,
as is evident from figures \ref{fig:evals}b and \ref{fig:evals}c, they
do not converge to well-defined limits. Instead, as the resolution
{$N^2$} is refined, the purely imaginary eigenvalues fill an expanding
subinterval of the imaginary axis and they do so ever more densely.
{We thus interpret them as representing points in the essential
  spectrum $\sigess(\L)$, cf.~\eqref{eq:sigess}, that do not belong to
  the point spectrum $\sigma_p(\L)$}.
\begin{figure}
  \centering
  \begin{subfigure}{0.45\textwidth}
    \includegraphics[width=\textwidth]{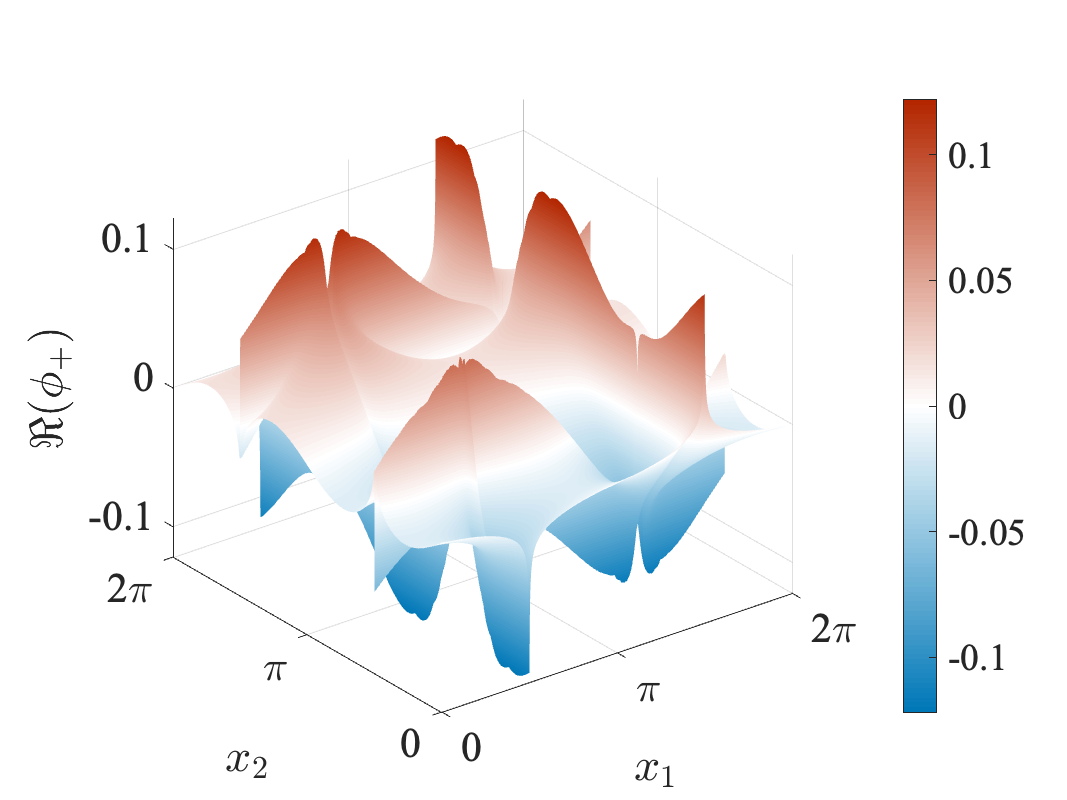}
    \caption{}
  \end{subfigure}
  \hfill
  \begin{subfigure}{0.45\textwidth}
    \includegraphics[width=\textwidth]{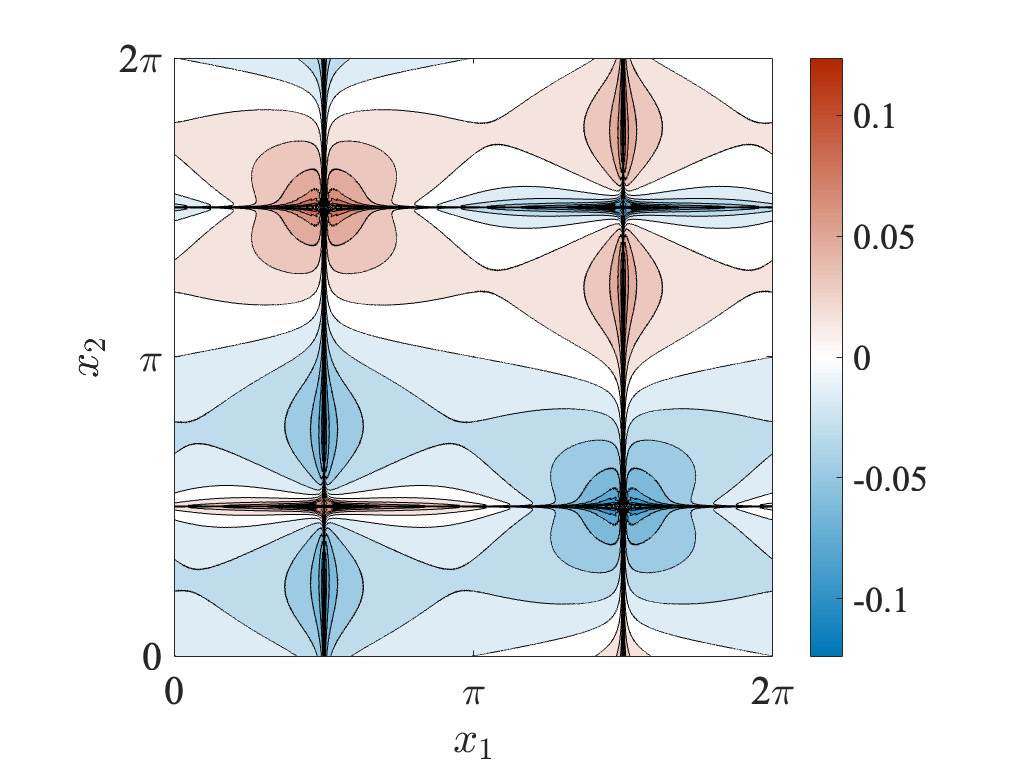}
    \caption{}
  \end{subfigure}
  
  \begin{subfigure}{0.45\textwidth}
    \includegraphics[width=\textwidth]{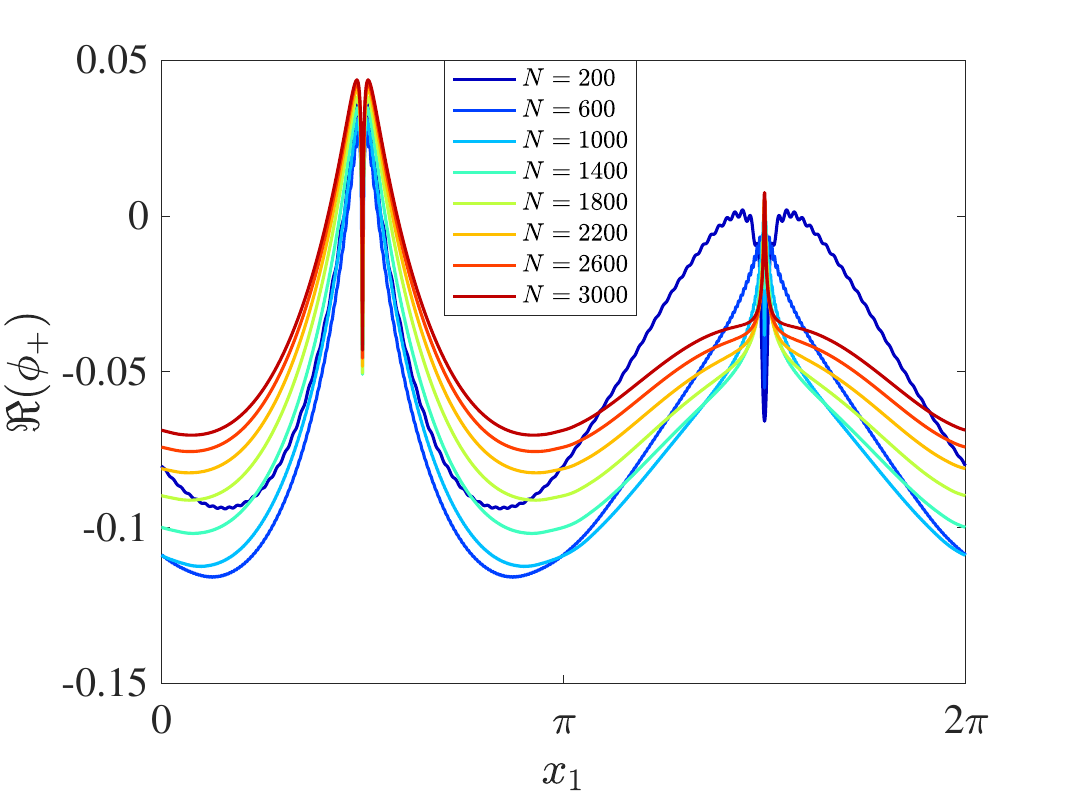}
    \caption{}
  \end{subfigure}
   \hfill
   \begin{subfigure}{0.45\textwidth}
    \includegraphics[width=\textwidth]{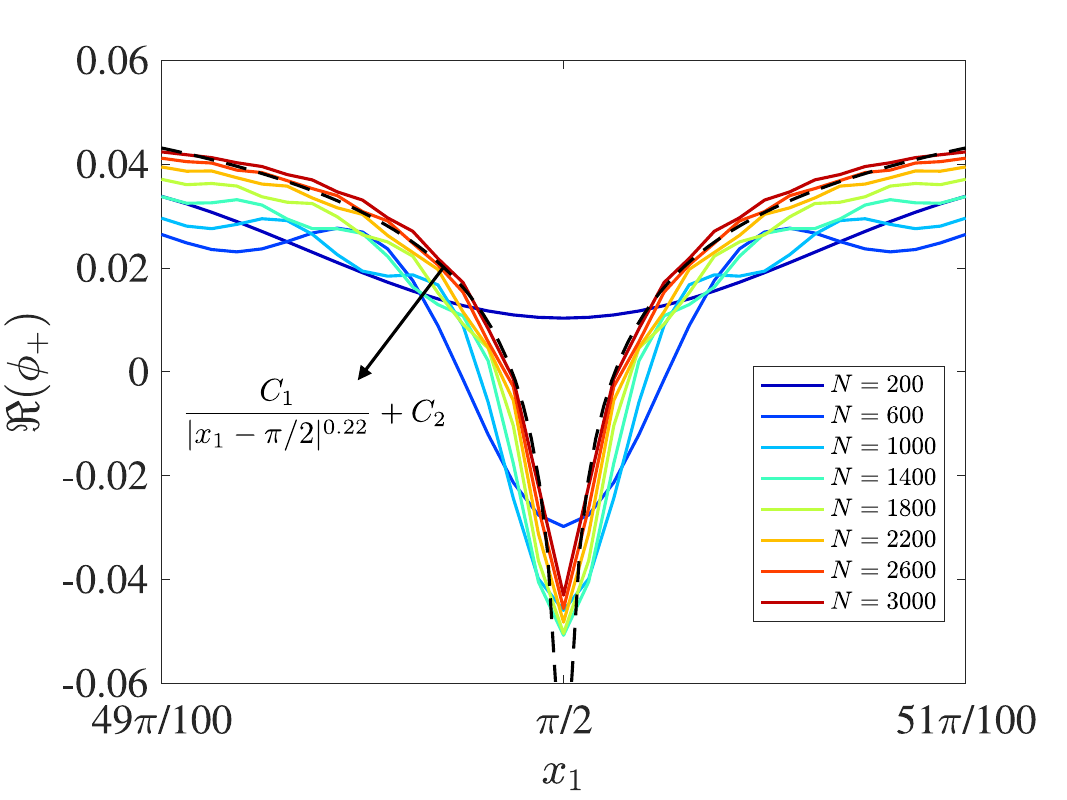}
    \caption{}
  \end{subfigure}
   \hfill
  \begin{subfigure}{0.45\textwidth}
    \includegraphics[width=\textwidth]{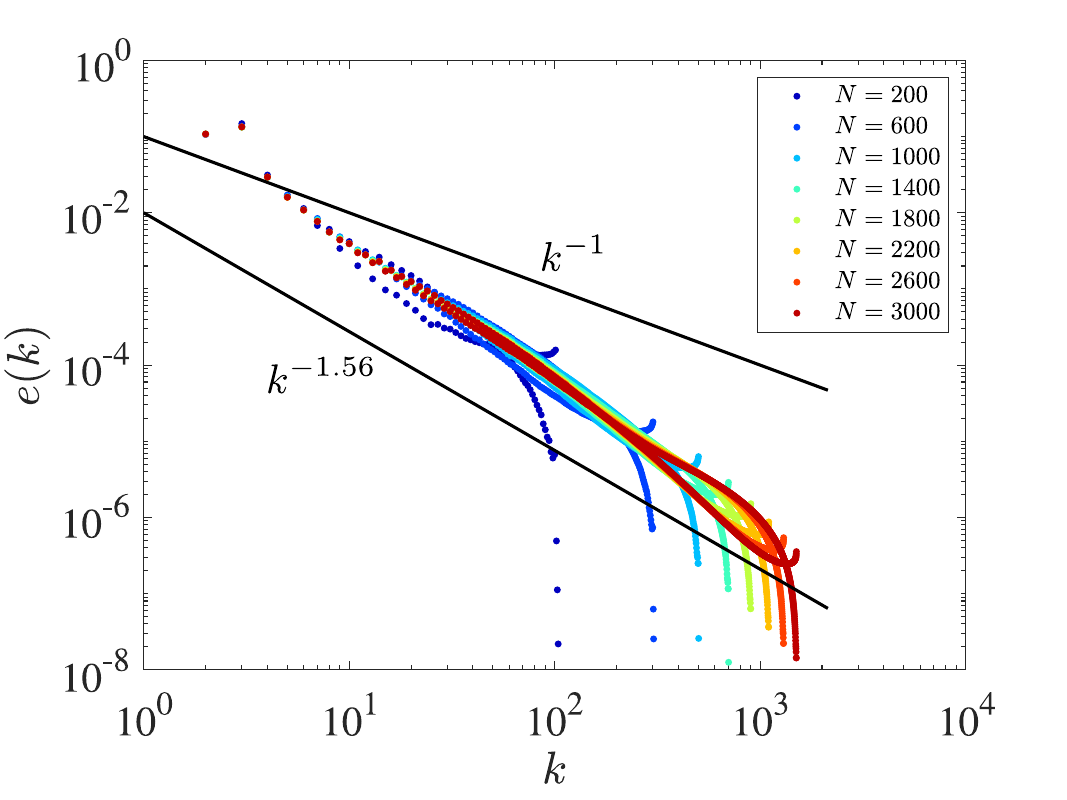}
    \caption{}
  \end{subfigure}
  \caption{(a) The surface plot and (b) the contour plot of the real
    part {$\Re\left(\phi_+^{3000}\right)$} of the eigenfunction
    corresponding to the eigenvalue {$\lambda_+^{3000}$} as a
    function of $x_1$ and $x_2$. (c) The cross-section
    {$\Re\left[\phi_+^N(x_1,\pi/2)\right]$} as a function of
    $x_1$ for different resolutions $N^2$ {with (d) showing a
      magnification of the neighbourhood of the stagnation point
      $(\pi/2,\pi/2)$; the dashed line represents the function
        $\frac{C_1}{|x_1-\pi/2|^{0.22}} + C_2$ with some $C_1,C_2 >
        0$, cf.~\eqref{eq:phixs}.} (e) The energy spectra
    \eqref{eq:ek} of the eigenfunctions $\phi_+^N$ for different
    resolutions {with the straight lines representing the
      power-law relations with indicated exponents. }}
\label{fig:eigfun}
\end{figure}

\subsubsection{Modal growth}
\label{sec:modal_growth}

\begin{figure}
  \centering
  \begin{subfigure}{0.45\textwidth}
    \includegraphics[width=\textwidth]{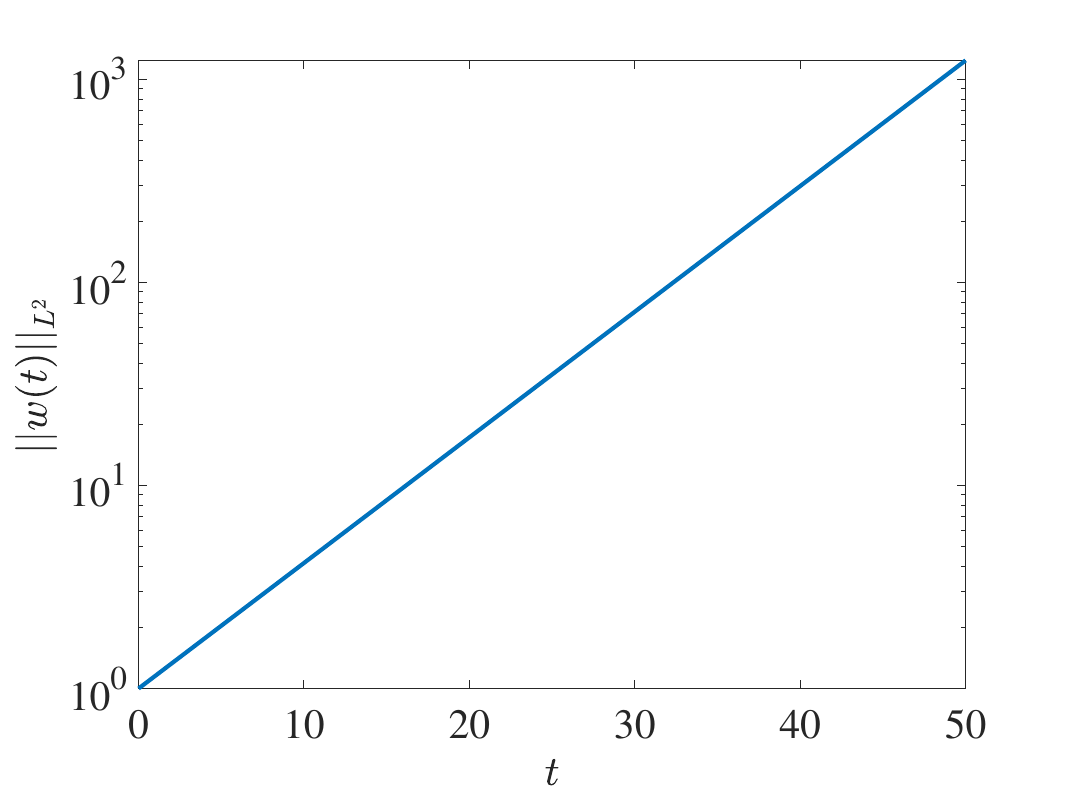}
    \caption{}
  \end{subfigure}
  \hfill
  \begin{subfigure}{0.45\textwidth}
    \includegraphics[width=\textwidth]{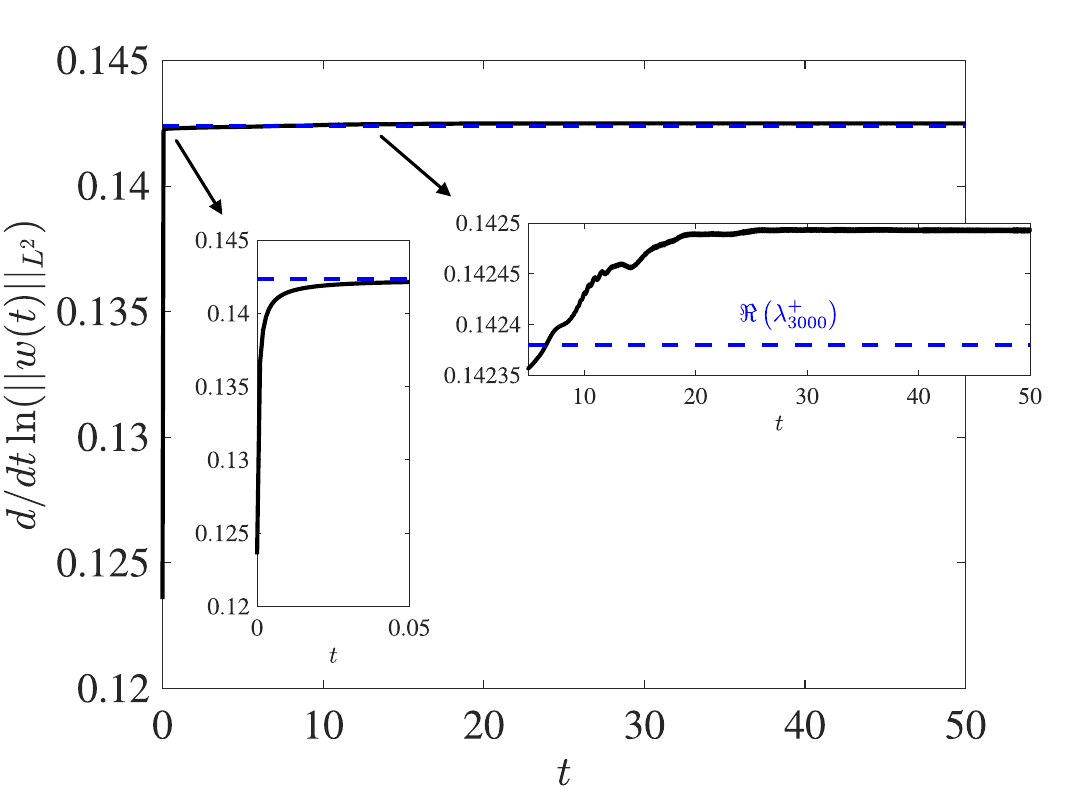}
    \caption{}
  \end{subfigure}
  \begin{subfigure}{0.45\textwidth}
    \includegraphics[width=\textwidth]{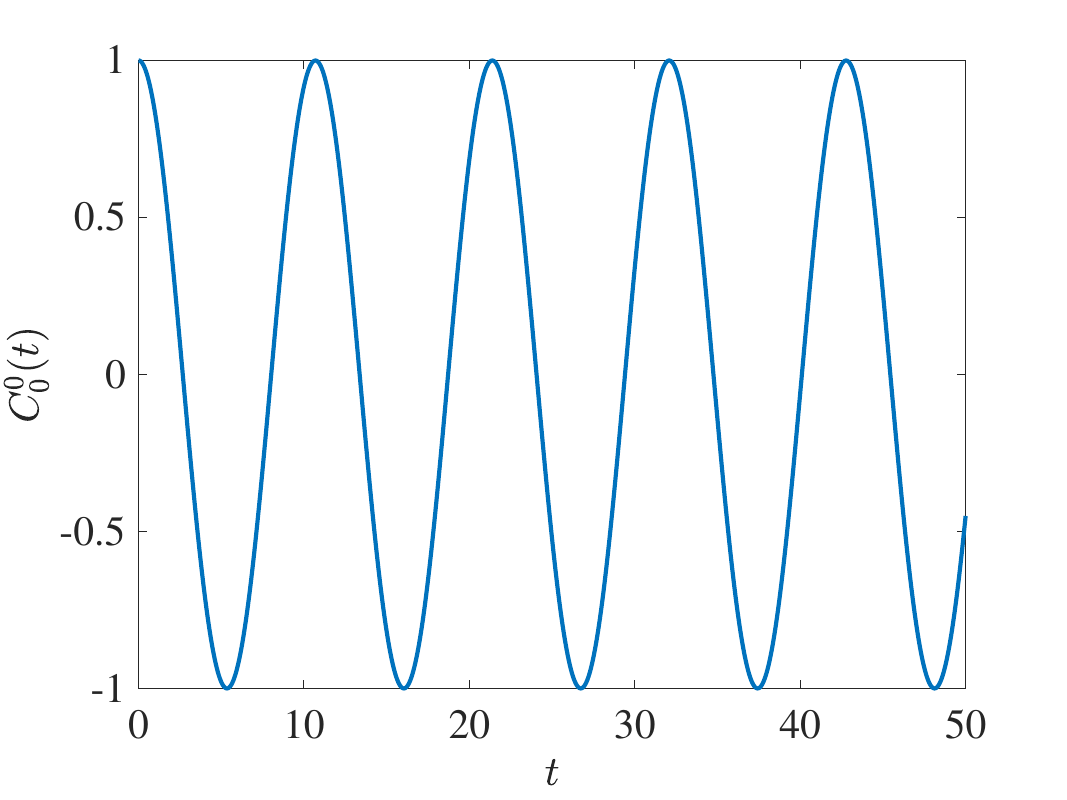}
    \caption{}
  \end{subfigure}
  \caption{The time-dependence of (a) the norm {$\| w(t)
      \|_{L^2}$}, (b) the growth rate {$(d/dt) \ln(\| w(t)
      \|_{L^2})$} and (c) the
    autocorrelation {$\C_{0}^{0}(t)$} corresponding to the solution of the linear
    problem \eqref{eq:vort:linear} with the initial condition given by
    the eigenfunction {$\Re\left(\phi_+^{3000}\right)$}. The dashed
    horizontal line in panel (b) corresponds to
    {$\Re\left(\lambda_+^{3000}\right)$, cf.~figure \ref{fig:evals}b.}}
  \label{fig:growth0}
\end{figure}

We now analyze the eigenfunction {$\phi_+^{3000}$} corresponding
to the eigenvalue {${\lambda_+^{3000}}$,} and its real part is
shown as a surface plot in figure \ref{fig:eigfun}a and as a contour
plot in figure \ref{fig:eigfun}b.  {We observe that
  $\Re\left[\phi_+^{3000}(x_1,x_2)\right]$ is an odd function and is
  also symmetric with respect to the lines $x_1 = \pi/2$,
  $x_1 = 3\pi/2$, $x_2 = \pi/2$, and $x_2 = 3\pi/2$; this is also true
  for $\Im\left[\phi_+^{3000}(x_1,x_2)\right]$ {and holds for
    the eigenfunctions obtained using lower resolutions as
    well. Therefore, these eigenfunctions satisfy the relations}
\begin{equation}
\begin{gathered}
\phi_+^N(-x_1, -x_2) = -\phi_+^N(x_1, x_2),\\
\phi_+^N(\zeta - x_1, x_2) = \phi_+^N(\zeta + x_1, x_2), \quad
\phi_+^N(x_1, \zeta - x_2) = \phi_+^N(x_1, \zeta + x_2),  \quad \zeta
= \frac{\pi}{2}, \frac{3 \pi}{2}.
\end{gathered}
\label{eq:eigsym}
\end{equation}
}

{We observe that $\Re\left[\phi_+^N\right]$ is localized near
  the hyperbolic stagnation points of the equilibrium \eqref{eq:TG}
  and to better understand the behaviour of the eigenfunction, in
  figure \ref{fig:eigfun}c we show
  $\Re\left[\phi_+^N(x_1,\pi/2)\right]$ as a function of $x_1$ for
  different resolutions $N^2$. In other words, this figure shows the
  cross-sections of the eigenfunction along the heteroclinic orbit
  connecting the hyperbolic stagnation points $(\pi/2, \pi/2)$ and
  $(\pi/2, 3\pi/2)$, with a magnification of the neighbourhood of the
  former shown in figure \ref{fig:eigfun}d.}

  {We now proceed to characterize the regularity of the eigenfunction
  $\phi_+$ more precisely. It is evident from figures
  \ref{fig:eigfun}a,c,d that $\phi_+$ is $L^2$-integrable and
  therefore we can refer to part (i) of Theorem \ref{thm:Lin},
  cf.~\eqref{eq:p*}, from which we conclude that
  $\phi_+ \in W^{1,p^*}(\TT^2)$ with
  $p^* = 1 / (1 - \Re(\lambda_+)) \approx 1.16$. So that we can
  compare this prediction with the regularity of the numerical
  approximations $\phi_+^N$ of the eigenfunction, we invoke the Sobolev
  embedding $W^{1,p^*} \hookrightarrow H^m$ with
  $1/p^* - 1/n = 1/2 - m / n$ \citep{af05} for the spatial dimension
  $n = 2$, which allows us to conclude that $s \approx 0.28$, such
  that $\phi_+ \in H^{0.28}_0(\TT^2)$. Since the regularity of a
  function, understood as the number of well-behaved derivatives, is
  encoded in the rate of decay of its Fourier coefficients as
  $k \rightarrow \infty$ \citep{n13}, we consider the energy spectra
  of the numerically computed eigenfunctions
\begin{equation}
e(k) := \frac{1}{2}\sum_{k \leq |\bds j| < k+1} \left|\widehat\phi_{\bds j}^N\right|^2, \qquad k \in \mbb N,
\label{eq:ek}
\end{equation}
where $\widehat\phi_{\bds j}^N$ are the Fourier coefficients of the
approximations $\phi_+^N$ obtained with different resolutions $N^2$.
This is an approach which has had a long tradition in fluid mechanics
\citep{bmonmu83,b91}. For a function to be $L^2$-integrable, its
energy spectrum \eqref{eq:ek} needs to vanish no slower than
$\mathcal{O}(k^{-1})$.  The energy spectra of the eigenfunctions
$\phi_+^N$ approximated with different resolutions $N^2$ are shown in
figure \ref{fig:eigfun}e. We see that, as the resolution is refined,
the energy spectrum decays approximately as $e(k) =
\mathcal{O}(k^{-1.56})$, which is consistent with $\phi_+ \in
H^{0.28}_0(\TT^2)$ predicted by Theorem \ref{thm:Lin}, demonstrating the
sharpness of this result. As a result of the symmetry of the
eigenfunction stated in \eqref{eq:eigsym}, we have $\widehat\phi_{j_1,
  j_2} = 0$ when $j_1 + j_2$ is even. We also observe that for
  each resolution, the energy spectrum splits into two branches, an
  effect that becomes more evident when the energy spectrum
  \eqref{eq:ek} is redefined to depend on the 1-norm of the wavevector
  $\bds j$, i.e., on $||\bds j||_1 := |j_1| + |j_2|$, rather than on $
  |\bds j|$. In such case $e(k)$ is at the level of round-off errors
  when $k$ is even.}

{To better understand the structure of the eigenfunction $\phi_+$ in
the neighbourhood of the hyperbolic stagnation point $\x_s = (\pi/2, \pi/2)$, we will
represent it locally in terms of the following asymptotic ansatz
\begin{equation}
\phi_+ \sim \frac{1}{|\xi_1|^\alpha |\xi_2|^\beta}, \qquad \text{where} \quad
\xi_i = x_i - \frac{\pi}{2}, \quad i=1,2,
\label{eq:phixs}
\end{equation}
reflecting the fact that the singularity in $\phi_+$ occurs along the
heteroclinic orbits of \eqref{eq:TG}. We then have $\partial_{\xi_1}^m
\phi_+ \sim 1 / \left( |\xi_1|^{\alpha+m} |\xi_2|^\beta \right)$ and
$\partial_{\xi_2}^m \phi_+ \sim 1 / \left( |\xi_1|^{\alpha}
  |\xi_2|^{\beta+m} \right)$, where $\partial_{\xi_i}^m$ is a partial
derivative of fractional order $m$. So that $\phi_+ \in
H^{0.28}_0(\TT^2)$, these expressions need to be square-integrable which
necessitates $2(\alpha + m) < 1$ and $2(\beta + m) < 1$. Therefore, we
arrive at $\alpha, \beta < 1/2 - m \approx 0.22$. In figure
\ref{fig:eigfun}d, we also show the function
  $\frac{C_1}{|x_1-\pi/2|^{0.22}} + C_2$ for some $C_1,C_2 > 0$ and
  conclude that it accurately represents the behavior of the
  eigenfunction $\Re\left[\phi_+^N(x_1,\pi/2)\right]$ near the point
  $x_1 = \pi/2$ as the resolution is refined.  This further confirms
  that $\phi_+$ is not continuous at the hyperbolic stagnation points.}

{We now comment on the rate of convergence of our numerical
approximations $\phi_+^N$ as the resolution is refined. Since the
eigenfunction is not smooth, we cannot expect the spectral approach
\eqref{eq:varphipsi}--\eqref{eq:evpN} to converge exponentially
fast. In fact, based on the standard convergence theory of spectral
methods \citep{Canuto1993book,ShenTangWang2011}, we have
\begin{equation} 
\left|\left|\phi_+^N - \phi_+\right|\right|_{L^2} \sim
\mathcal{O}\left(N^{-s}\right) \qquad \text{with} \quad s \approx 0.28.
\label{eq:Err1}
\end{equation}
To verify this prediction when the exact solution $\phi_+$ is not
available, we consider the quantity
\begin{equation} 
\frac{\left|\left|\phi_+^{N+\Delta N} -  \phi_+^N\right|\right|_{L^2}}{\Delta N} \sim \mathcal{O}\left(N^{-1-s}\right)
\label{eq:Err2}
\end{equation}
which can be viewed as an approximation of the ``derivative'' of
\eqref{eq:Err1} with respect to $N$. The left-hand side can be
evaluated using the approximations $\phi_+^N$ obtained at different
resolutions. Doing this for $N = 600, 800, \ldots, 3000$ and
$\Delta N = 200$, and performing a least-squares fit for the expression
on the right-hand side, we obtain $s = 0.29$ which confirms that our
approximations converge at an algebraic rate close to the
theoretical prediction in \eqref{eq:Err1}.
}

We now move on to analyze the growth of the solution $w(t)$ of the
linear system \eqref{eq:vort:linear} with the initial condition given
in terms of the unstable eigenfunction discussed above, i.e., with
${w(0) = \Re\left(\phi_+^{3000}\right)}$.  {The linear
  system is approximated using the spatial resolution
    $N^2 = 3200^2$ and the time step $\Delta t = 2^{-10}$.} The
dependence of the norm {$\|w(t) \|_{L^2}$} on time $t$ is shown
in figure \ref{fig:growth0}a revealing the expected exponential
growth.  The corresponding exponential growth rate
{$(d/dt)\ln (\| w(t) \|_{L^2}) $}, which is equal to the slope of
the curve in figure \ref{fig:growth0}a, is shown in figure
\ref{fig:growth0}b.  We observe that after a brief initial transient,
the growth rate settles at {0.1425}, which is to within less than
{0.08\%} equal to {$\Re\left(\lambda_+^{3000}\right)$}.
Finally, we consider the (normalized) autocorrelation function
\begin{equation}
\C^m_{\tau}(t) := \frac{\left\langle w(\tau), w(t) \right\rangle_{H^m}}{\| w(\tau) \|_{H^m} \| w(t) \|_{H^m}}, 
\qquad t, \tau \ge 0, \quad m \in \ZZ,
\label{eq:C}
\end{equation}
which in figure \ref{fig:growth0}c is shown for {$m = 0$ and
$\tau = 0$}. The harmonic behavior of the autocorrelation function
{$C^{0}_{0}(t)$} indicates that the solution $w(t)$ of the linear
system \eqref{eq:vort:linear} is at all times $t \ge 0$ a linear
combination of $\Re\left(\phi_+\right)$ and
$\Im\left(\phi_+\right)$. The oscillation period $\Delta T$ of the
autocorrelation function is related to the imaginary part of the
eigenvalue $\lambda_+$ and can be approximated by
{$\Delta T \approx 2\pi / \Im\left( \lambda_+^{3000} \right) = 10.6945$},
which is consistent with the results presented in figure
  \ref{fig:growth0}c. The behavior observed in figures
\ref{fig:growth0}a--c is typical for the {\em modal} growth of a
perturbation in a linear problem \citep{Schmid2001book} and further
confirms that the eigenvalue $\lambda_+$ and the corresponding
  eigenfunction $\phi_+$ obtained by solving the discrete eigenvalue
problem \eqref{eq:evpN} are indeed good numerical approximations of
the ``true'' eigenvalue and eigenfunction of problem
  \eqref{eq:sig0}.

\subsubsection{Nonmodal growth}
\label{sec:nonmodal_growth}

\begin{figure}
  \centering
  \begin{subfigure}{0.48\textwidth}
\includegraphics[width=\textwidth]{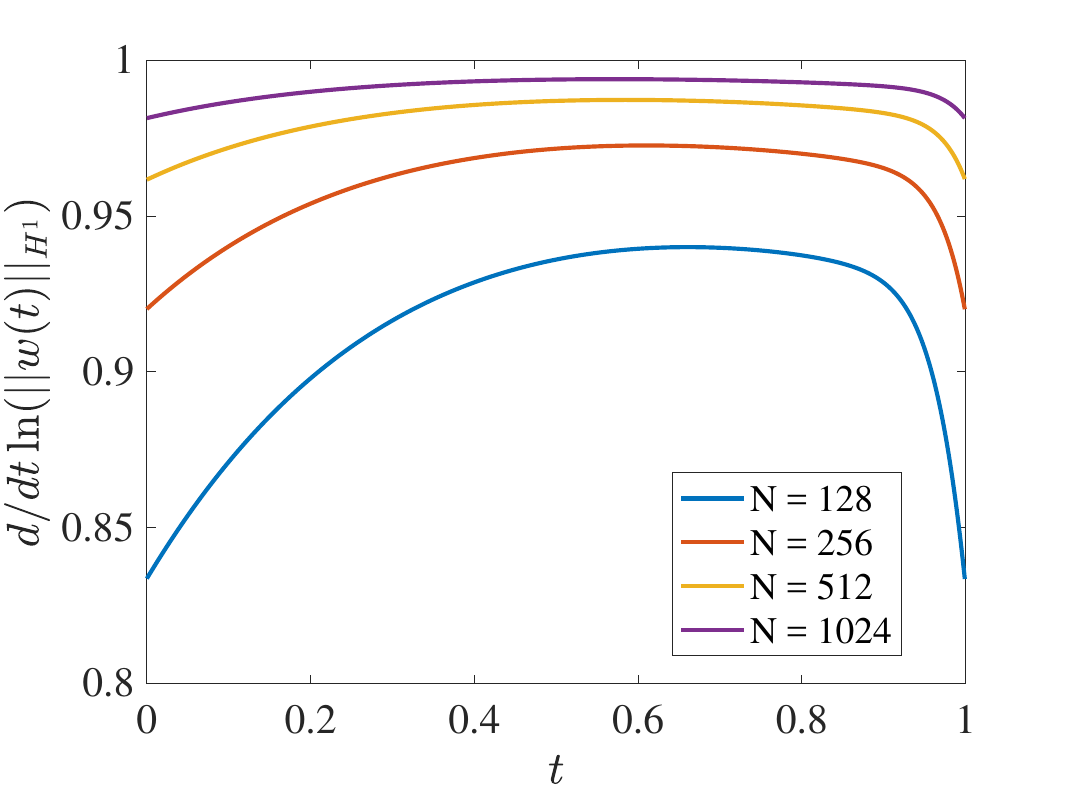}
\caption{}
\end{subfigure}
\hfill
 \begin{subfigure}{0.48\textwidth}
\includegraphics[width=\textwidth]{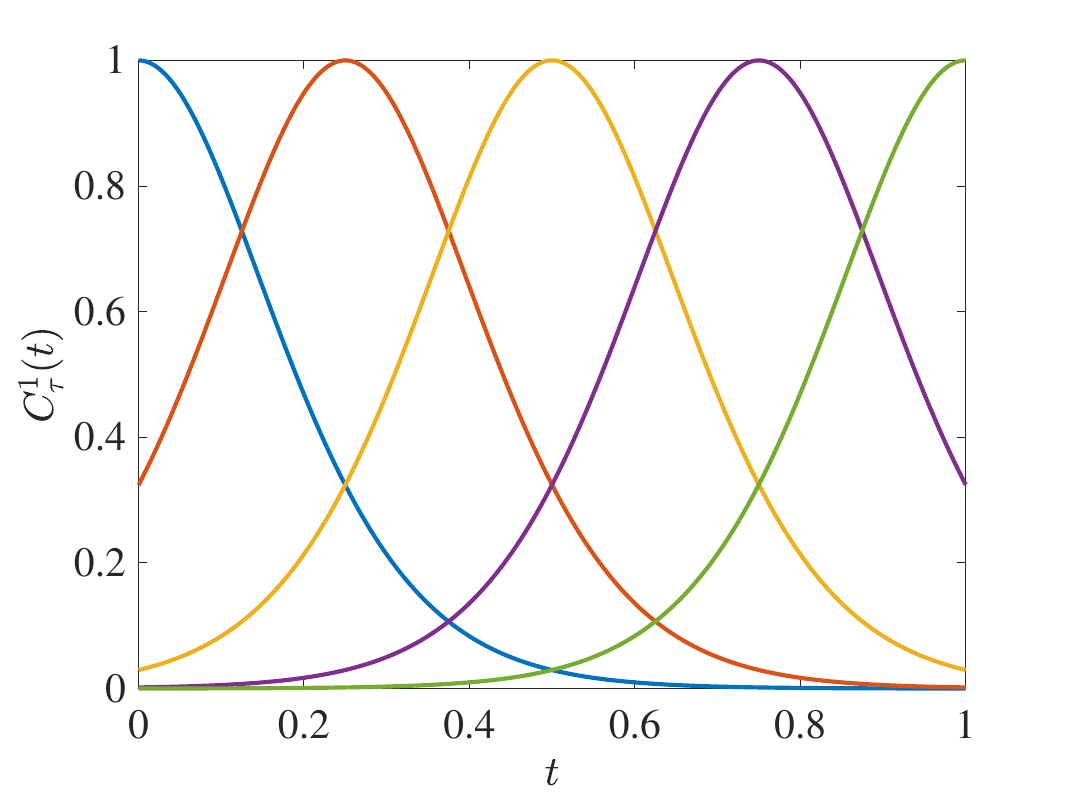}
\caption{}
\end{subfigure}
\caption{(a) The growth rate
  $(d/dt) \ln (\|w^N(t)\|_{H^{1}})$ versus $t$ for
  the optimal initial conditions $\wopt_0^N$ obtained by
  solving Problem \ref{prob:ess:growth} with $m = 1$ using increasing
  spatial resolutions {$N^2$}. (b) The time-dependence of the
  autocorrelation function $\C_\tau^{1}(t)$ corresponding to
  $N = 1024$ and {$\tau = 0, 0.25, 0.5, 0.75, 1$}.}
  \label{fig:s:1}
\end{figure}
\begin{figure}
  \centering
  \begin{subfigure}{0.48\textwidth}
\includegraphics[width=\textwidth]{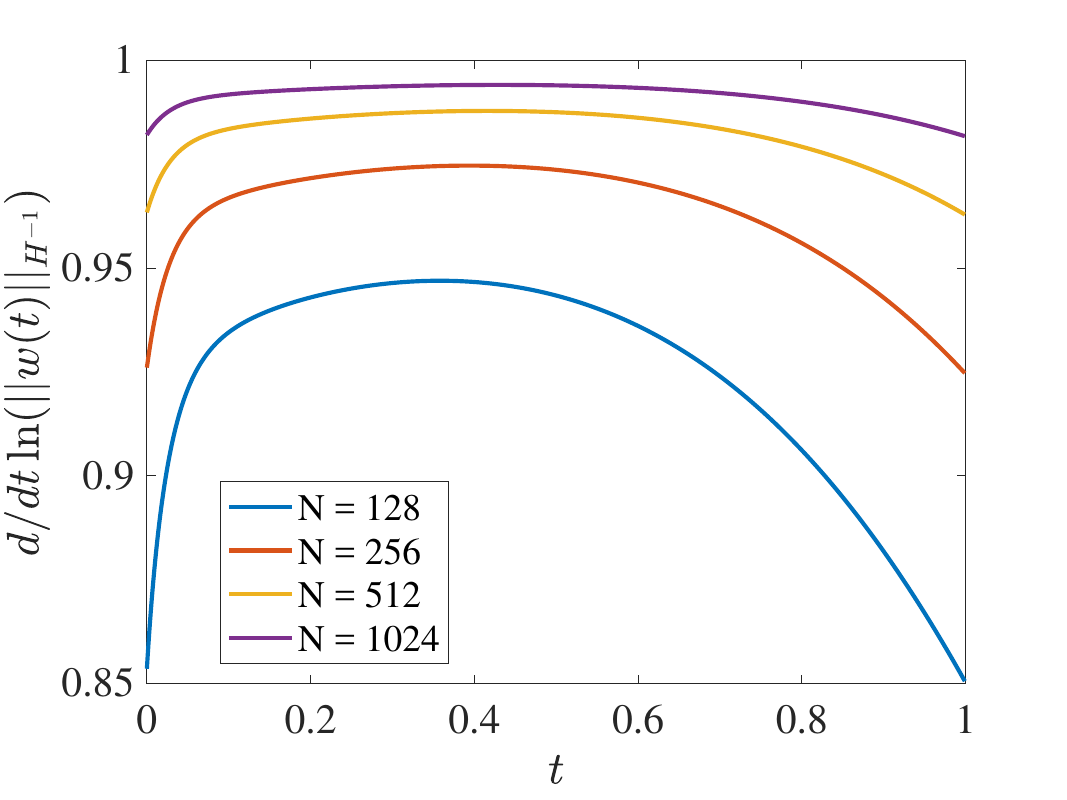}
\caption{}
\end{subfigure}
\hfill
 \begin{subfigure}{0.5\textwidth}
\includegraphics[width=\textwidth]{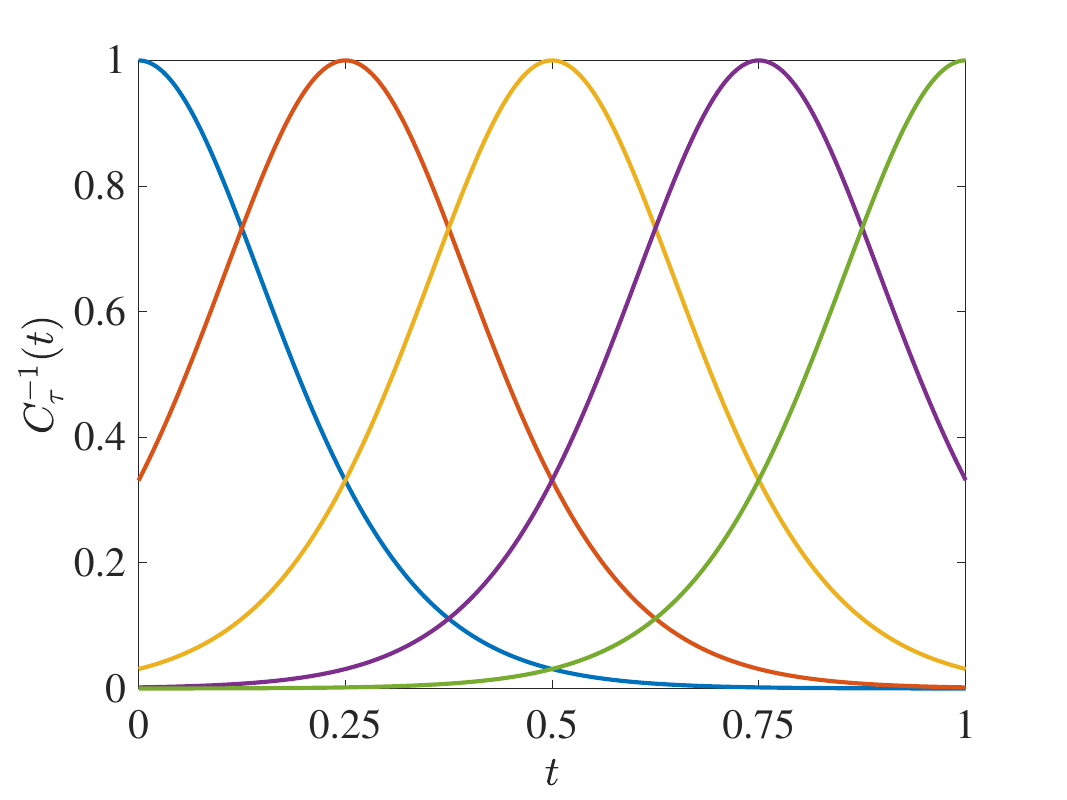}
\caption{}
\end{subfigure}
\caption{(a) The growth rate
  ${d/dt \ln (\|w^N(t)\|_{H^{-1}})}$ versus $t$ for
  the optimal initial conditions $\wopt_0^N$ obtained by
  solving Problem \ref{prob:ess:growth} with $m = 1$ using increasing
  spatial resolutions {$N^2$}. (b) The time-dependence of the
  autocorrelation function {$\C_\tau^{-1}(t)$} corresponding to
  $N = 1024$ and {$\tau = 0, 0.25, 0.5, 0.75, 1$}.}
  \label{fig:s:-1}
\end{figure}

In order to achieve the growth rate $\mu_{\max} = 1$ of the semigroup
$e^{t \L}$ predicted by the essential spectrum when $m = \pm 1$,
cf.~\eqref{eq:SMT}, we solve Problem \ref{prob:ess:growth}
with $m = \pm 1$ over a relatively short time window with $T = 1$ and
using increasing resolutions $N^2 = 128^2, 256^2, 512^2, 1024^2$. At
the lowest resolution $N^2 =128^2$, the initial guess $w^{(0)}
(x_1,x_2) = -\cos(x_2)$ is used in algorithm \eqref{eq:RCG}, and then,
for increasing resolutions, the optimal initial condition $\wopt_0^N$
obtained with the resolution $N^2$ is used as the initial guess in the
solution of the problem with the resolution $(2N)^2$.  In figures
\ref{fig:s:1}a and \ref{fig:s:-1}a we see that as the resolution $N^2$
is refined, the growth rate $(d/dt) \ln(\| w^N(t) \|_{H^{m}})$ with,
respectively, $m = 1$ and $m = -1$, approaches ${\mumax} = 1$ and is
sustained over an increasingly longer time. Thus, the optimal
flow evolutions found in this way indicate that the largest
  possible growth of the semigroup $e^{t \L}$ is associated, via the
  Spectral Mapping Theorem \eqref{eq:SMT}, with the essential spectrum
  \eqref{eq:sigess} of the operator $\L$. In other words, there are no
  eigenvalues {\em outside} the essential spectrum.

\begin{figure}
\centering
 \begin{subfigure}{0.5\textwidth}
\includegraphics[width=\textwidth]{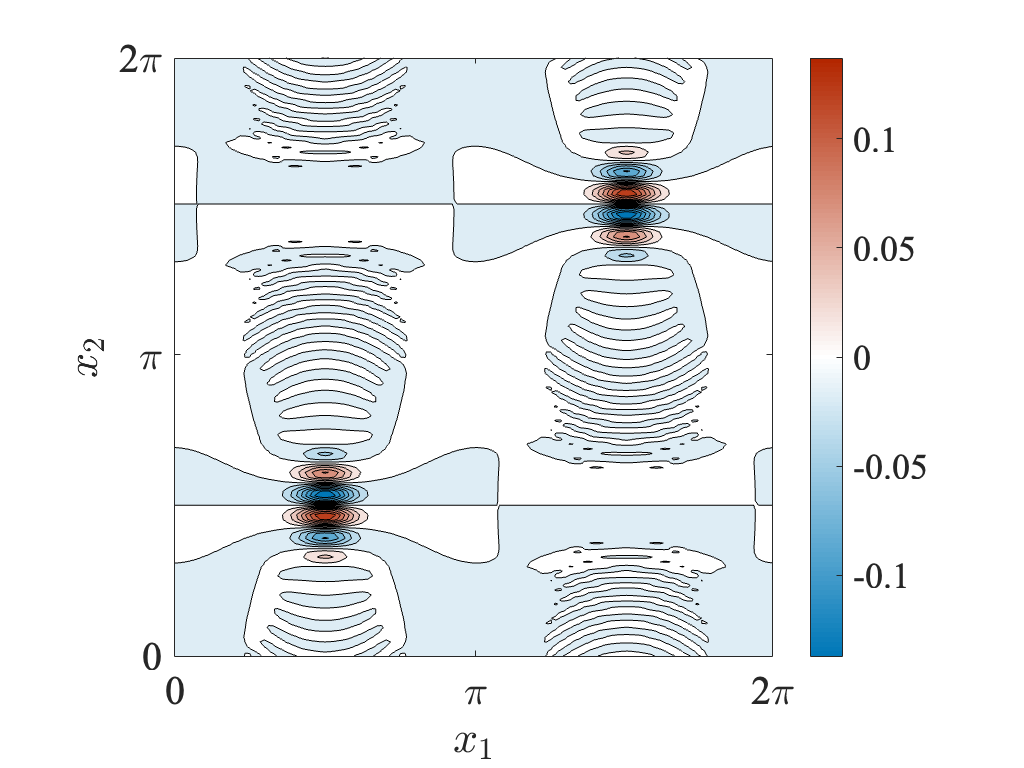}
\caption{}
\end{subfigure}

\begin{subfigure}{0.48\textwidth}
\includegraphics[width=\textwidth]{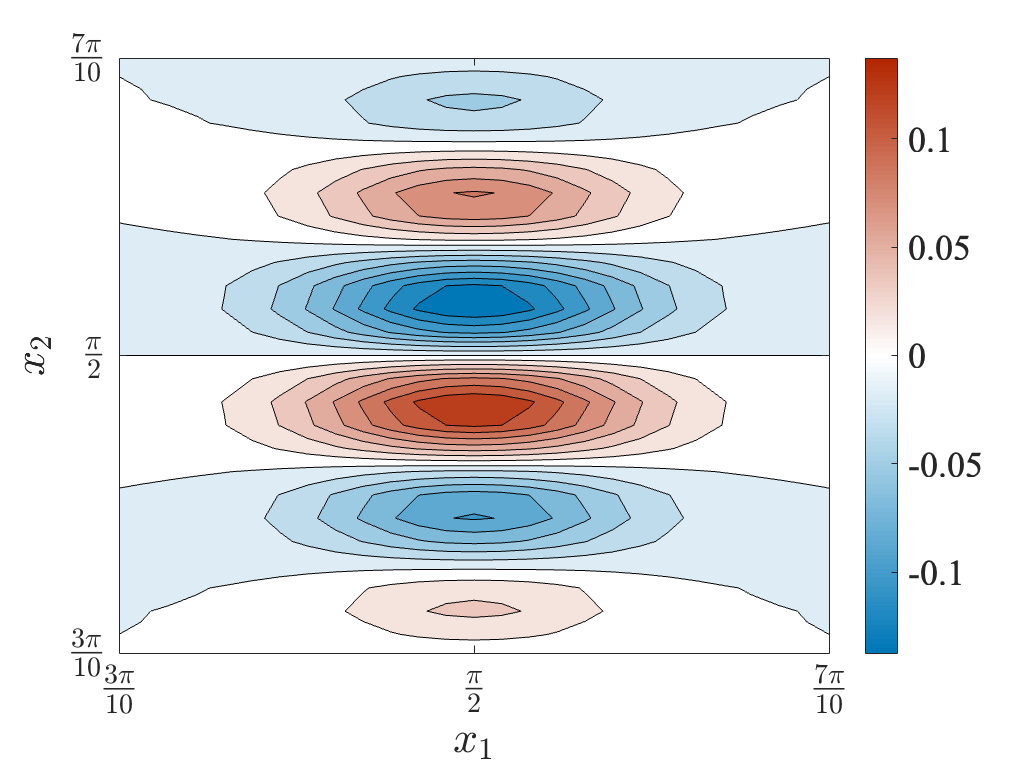}
\caption{}
\end{subfigure}
\begin{subfigure}{0.48\textwidth}
\includegraphics[width=\textwidth]{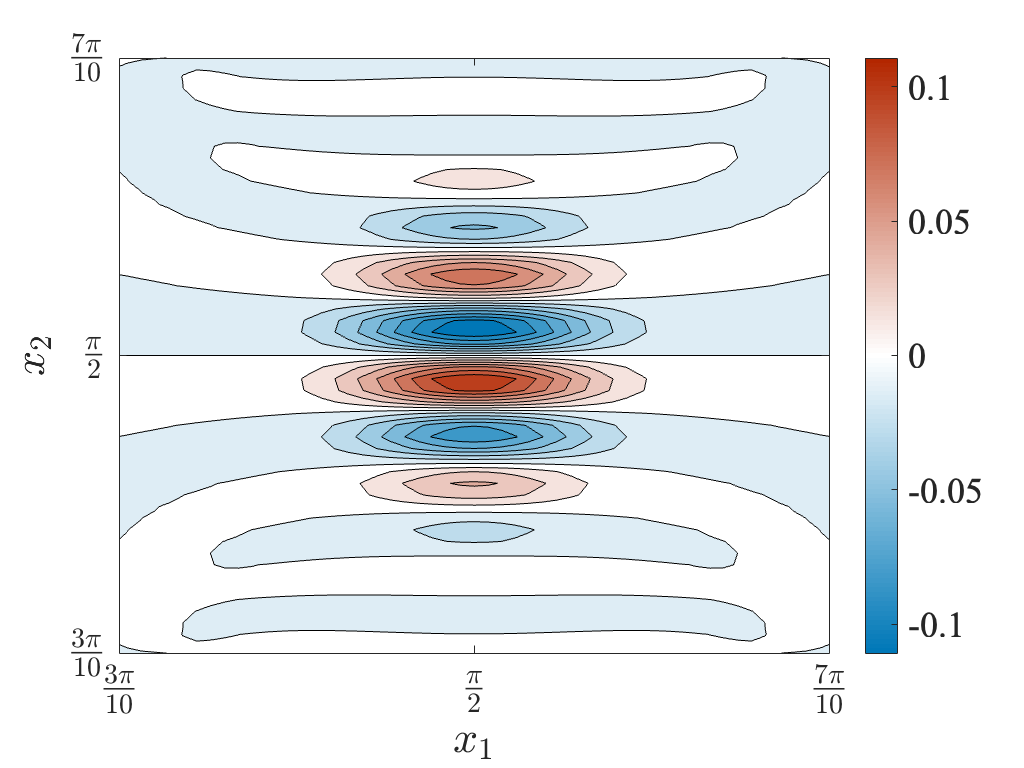}
\caption{}
\end{subfigure}

\begin{subfigure}{0.48\textwidth}
\includegraphics[width=\textwidth]{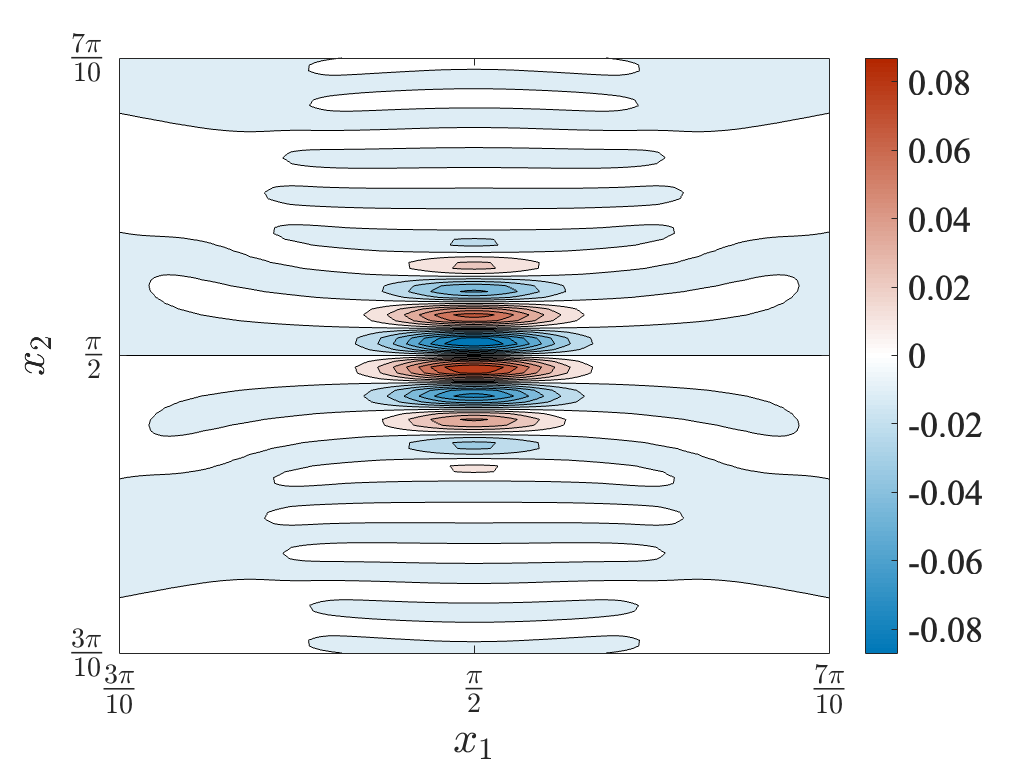}
\caption{}
\end{subfigure}
\begin{subfigure}{0.48\textwidth}
\includegraphics[width=\textwidth]{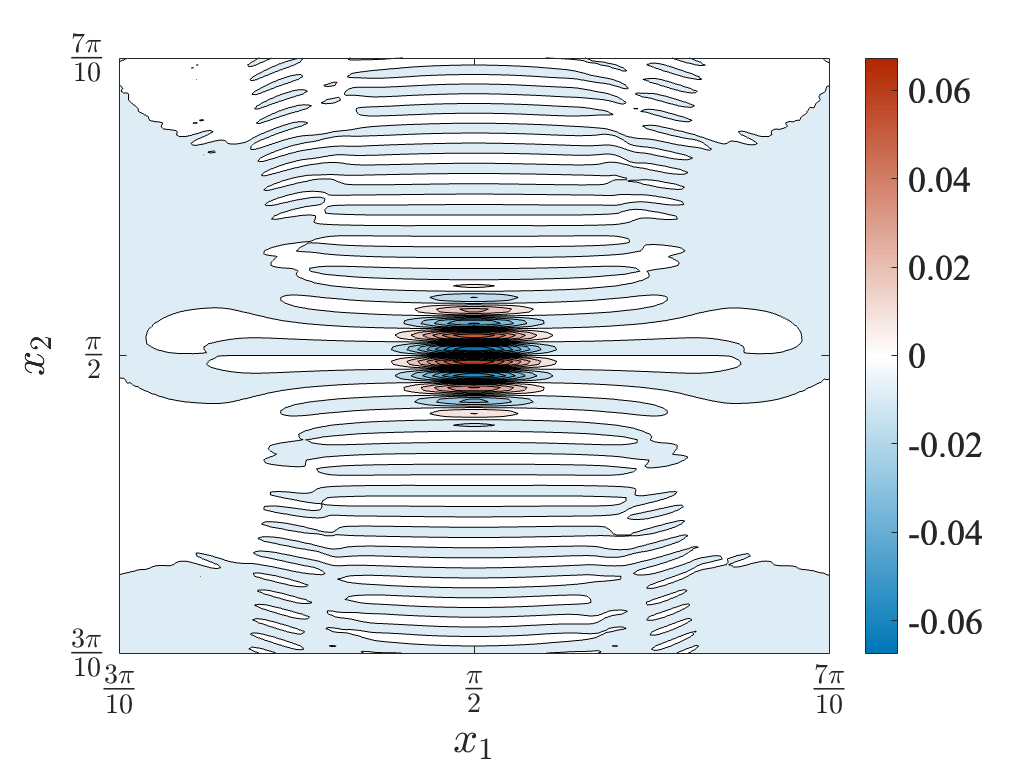}
\caption{}
\end{subfigure}

\caption{Contour plots of {the optimal initial conditions} obtained by
  solving Problem \ref{prob:ess:growth} with $m = 1$: (a)
  $\wopt_0^{128}$ is shown in the entire domain $\TT^2$ and (c--e)
  $\wopt_0^{N}$ are shown for $N = 128, 256, 512,1024$ near the
  hyperbolic stagnation point {$(\pi/2, \pi/2)$}. The time evolution
  of the flow corresponding to the initial condition $\wopt_0^{1024}$,
  cf.~panel (e), is shown in movie 1.}
  \label{fig:cont:s:1}
\end{figure}
\begin{figure}
\centering
 \begin{subfigure}{0.5\textwidth}
\includegraphics[width=\textwidth]{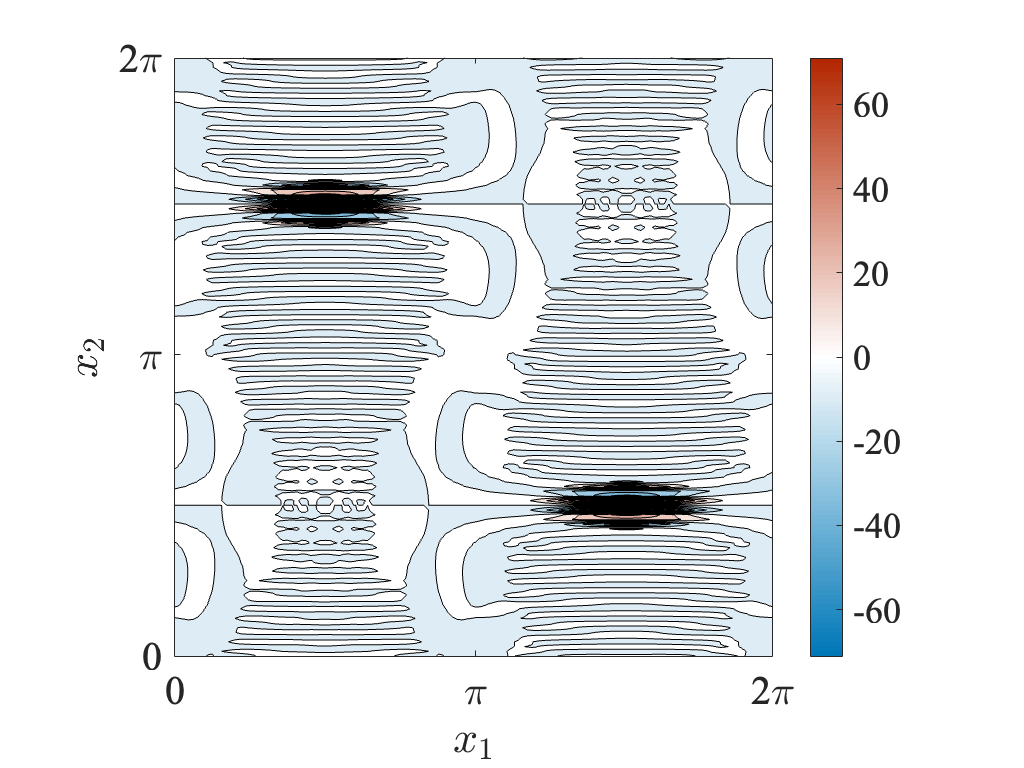}
\caption{}
\end{subfigure}

\begin{subfigure}{0.48\textwidth}
\includegraphics[width=\textwidth]{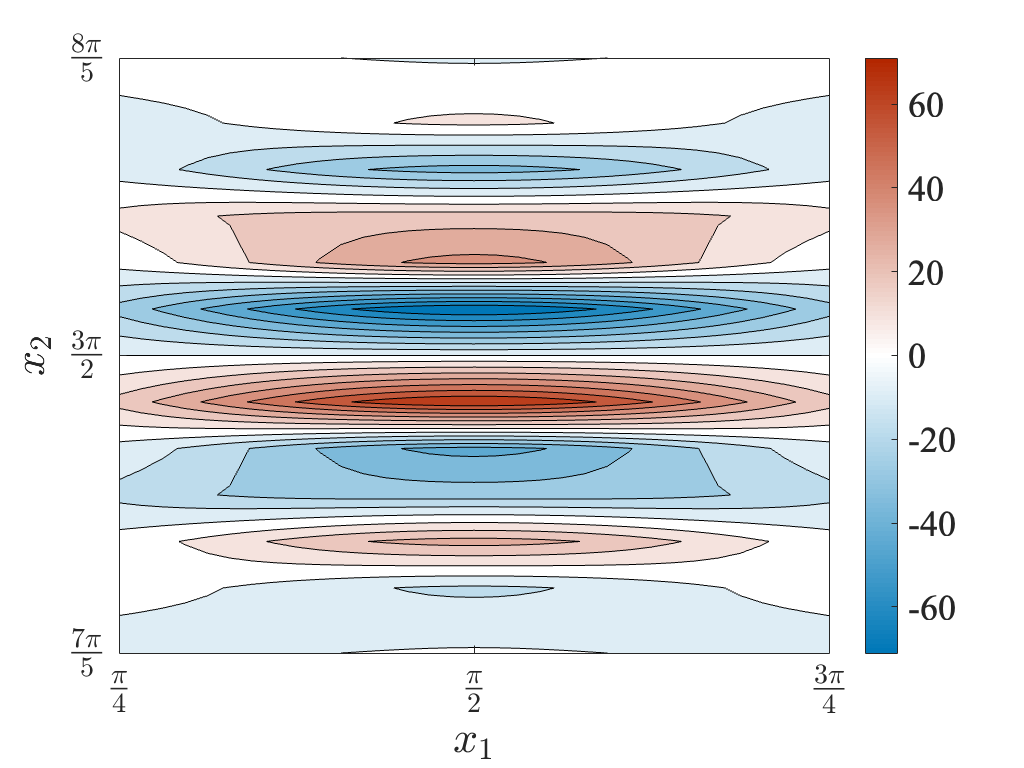}
\caption{}
\end{subfigure}
\begin{subfigure}{0.48\textwidth}
\includegraphics[width=\textwidth]{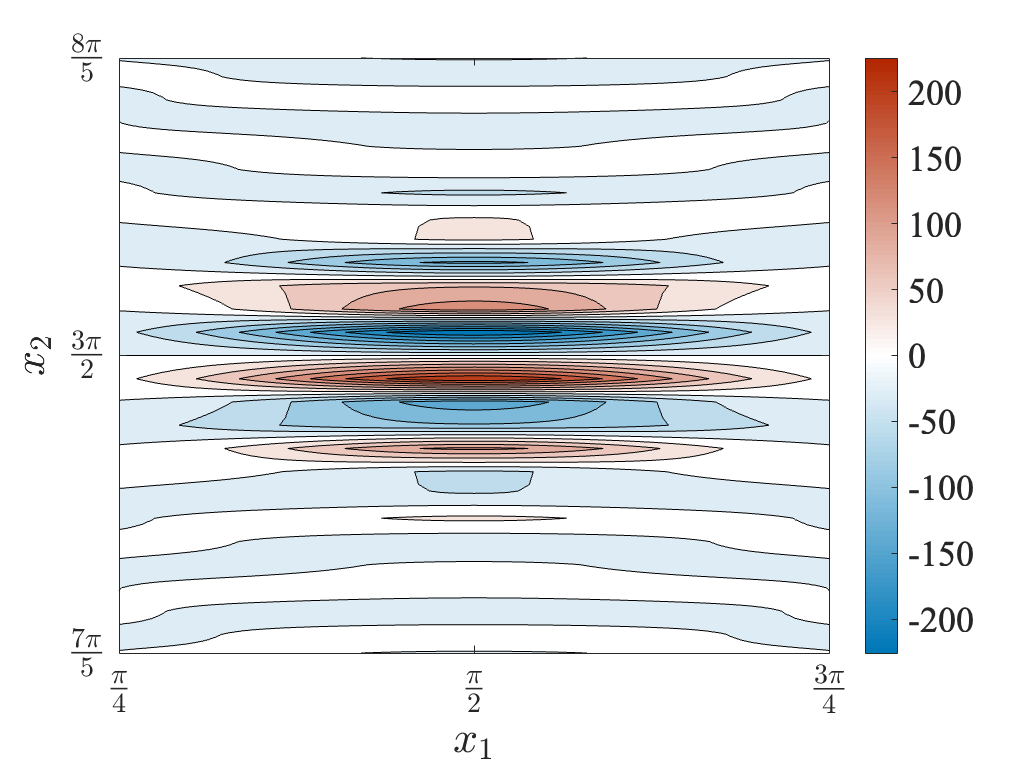}
\caption{}
\end{subfigure}

\begin{subfigure}{0.48\textwidth}
\includegraphics[width=\textwidth]{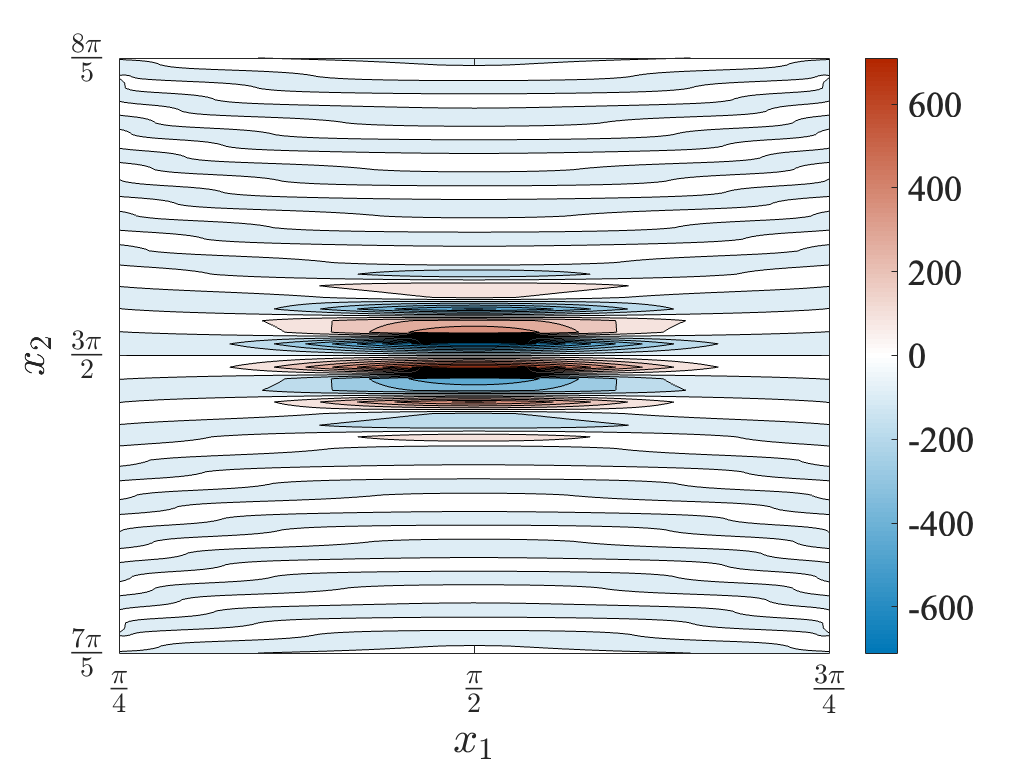}
\caption{}
\end{subfigure}
\begin{subfigure}{0.48\textwidth}
\includegraphics[width=\textwidth]{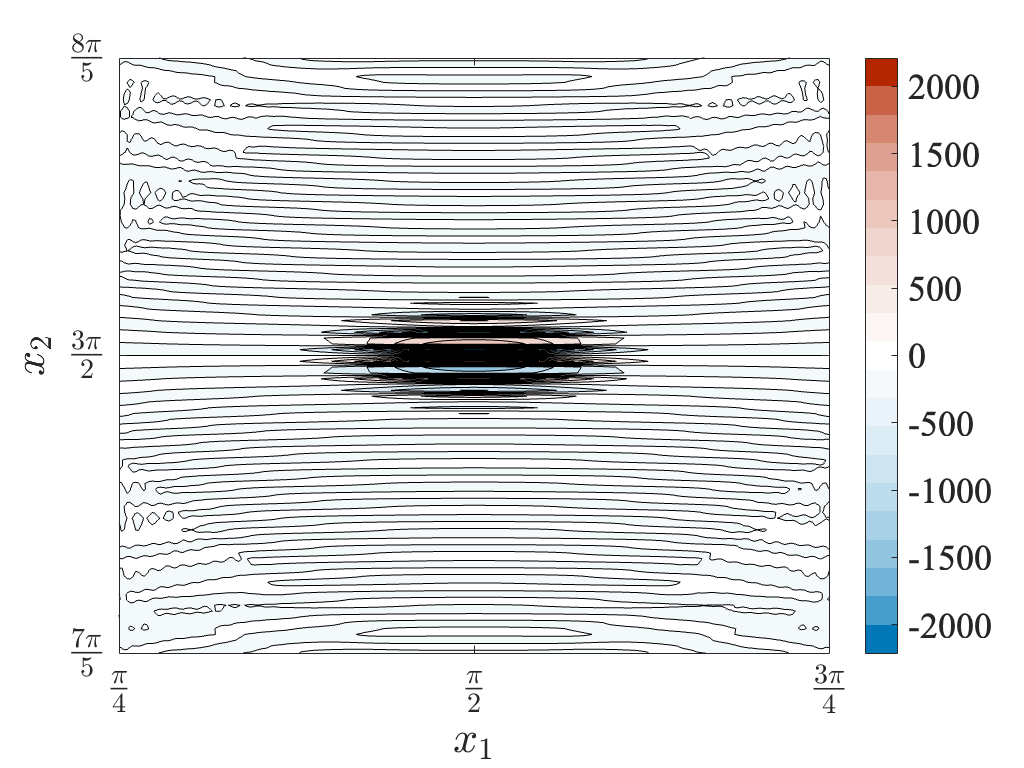}
\caption{}
\end{subfigure}

\caption{Contour plots of {the optimal initial conditions} obtained by
  solving Problem \ref{prob:ess:growth} with $m = - 1$: (a)
  $\wopt_0^{128}$ is shown in the entire domain $\TT^2$ and (c--e)
  $\wopt_0^{N}$ are shown for $N = 128, 256, 512,1024$ near the
  hyperbolic stagnation point $(\pi/2, 3\pi/2)$.  The time evolution
  of the flow corresponding to the initial condition $\wopt_0^{1024}$,
  cf.~panel (e), is shown in movie 2.}
  \label{fig:cont:s:-1}
\end{figure}

  The contour plots of the optimal initial conditions $\wopt_0^{128}$
  are shown in figures \ref{fig:cont:s:1}a and \ref{fig:cont:s:-1}a,
  respectively, for $m = 1$ and $m = -1$, where we see that {similarly
    to the ``true'' eigenfunction ${\phi_+}$, cf.~figure
    \ref{fig:eigfun}a--\ref{fig:eigfun}b}, these optimal initial
  conditions are also localized around the hyperbolic stagnation
  points of the equilibrium flow \eqref{eq:TG}. How these small-scale
  features are refined as the resolution ${N^2}$ increases is shown
  for $m = 1$ and $m = -1$ in figures \ref{fig:cont:s:1}b--e and
  \ref{fig:cont:s:-1}b--e which present magnifications of the
  neighbourhoods of the stagnation points $(\pi/2, \pi/2)$ and
  $(\pi/2, 3\pi/2)$, respectively. We see that, in contrast to the
  ``true'' eigenfunction $\phi_+$, the the optimal initial conditions
  $\wopt_0^N$ feature small-scale oscillations that become
  increasingly concentrated at the stagnation points as the resolution
  $N^2$ increases with the length scale of the oscillations restricted
  by the spatial resolution used.  When $m = 1$, these oscillations
  are localized along the stable manifolds and stretched along the
  unstable ones, cf.~figure \ref{fig:cont:s:1}b--e, and vice versa
  when $m = -1$, cf.~figure \ref{fig:cont:s:-1}b--e.  Unlike the
  sequence {$\{\phi_+^N\}$} which converges to the true eigenfunction
  $\phi_+$ as $N$ increases, {cf.~\eqref{eq:Err1},}  the
  sequence {$\{\wopt_0^N\}$} does not converge in a strong sense and
  this lack of compactness underpins the infinite-dimensional nature
  of the stability problem.


Finally, in figures \ref{fig:s:1}b and \ref{fig:s:-1}b we show the
autocorrelation function \eqref{eq:C}, respectively, for $m = 1$ and
$m = -1$ and $\tau = 0, 0.25, 0.5, 0.75, 1$. We see that the behavior
in these plots is fundamentally different from what {is} observed in
figure \ref{fig:growth0}c, in that here the autocorrelation function
decays quite rapidly, indicating that the solution $w(t)$ becomes
effectively decorrelated after about half time unit.  In other words,
there is no single growing mode and instead the evolution $w(t)$ moves
through a continuous family of essentially uncorrelated functions.
For this reason, we refer to the perturbation growth analyzed here as
``nonmodal''. The time evolution of the nonmodal perturbations
realizing the behavior shown in figures \ref{fig:s:1} and
\ref{fig:cont:s:1} for $m = 1$ and in figures \ref{fig:s:-1} and
\ref{fig:cont:s:-1} for $m = -1$ is visualized in movies 1 and 2 (they
are also available, respectively, at
\href{https://youtu.be/O8xM_1OvuHI}{https://youtu.be/O8xM\_1OvuHI} and
\href{https://youtu.be/jLgUvRKPZ7o}{https://youtu.be/jLgUvRKPZ7o}).


\subsection{Nonlinear {instability}}
\label{sec:nonlinear_growth}

\begin{figure}
  \centering
  \begin{subfigure}{0.48\textwidth}
    \includegraphics[width=\textwidth]{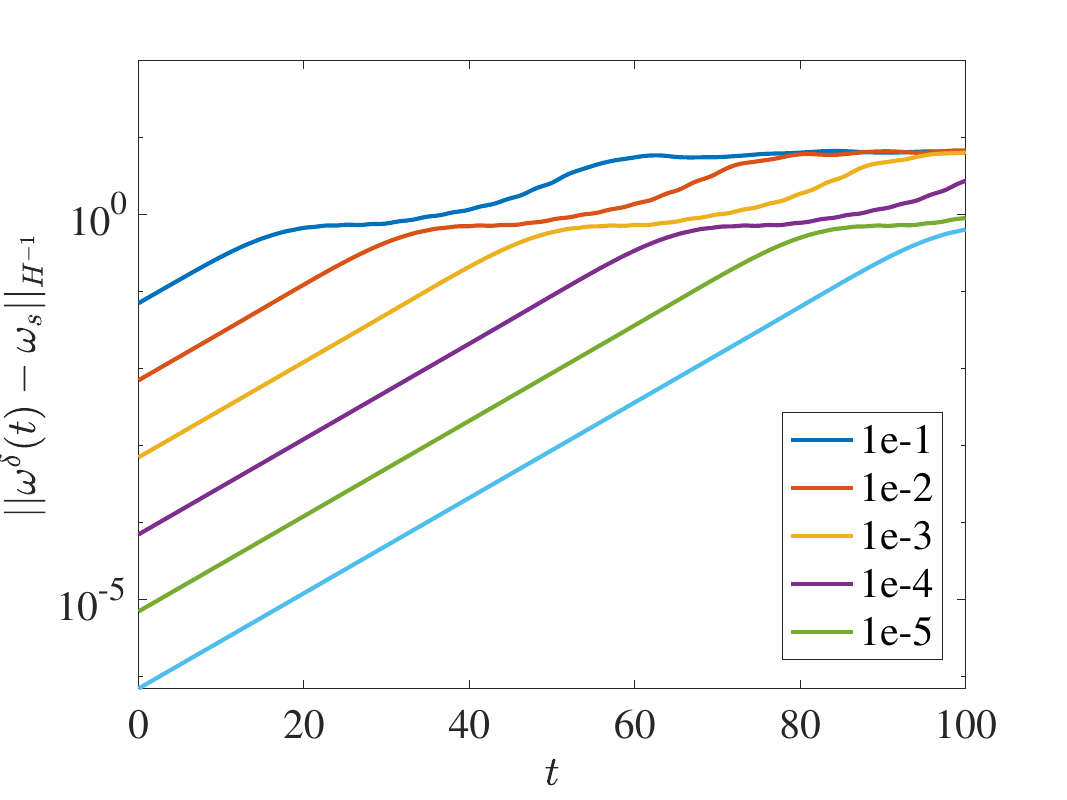}
    \caption{}
  \end{subfigure}
  \hfill
  \begin{subfigure}{0.48\textwidth}
    \includegraphics[width=\textwidth]{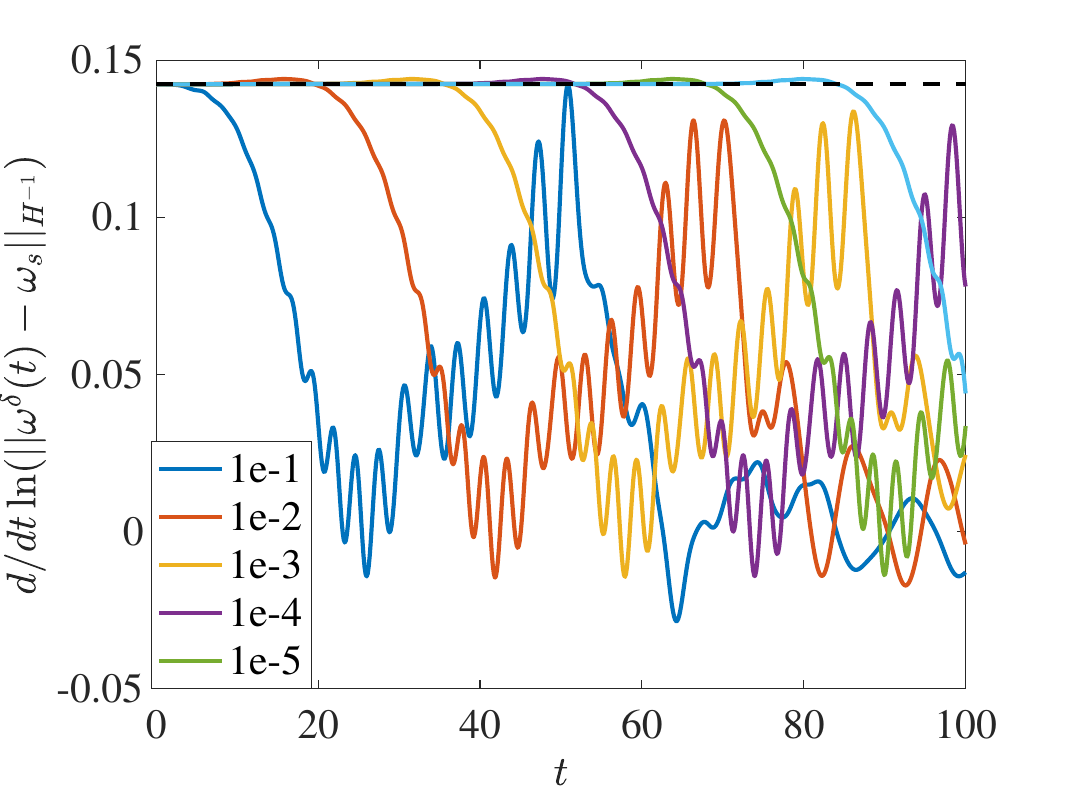}
    \caption{}
  \end{subfigure}
  
  \begin{subfigure}{0.48\textwidth}
    \includegraphics[width=\textwidth]{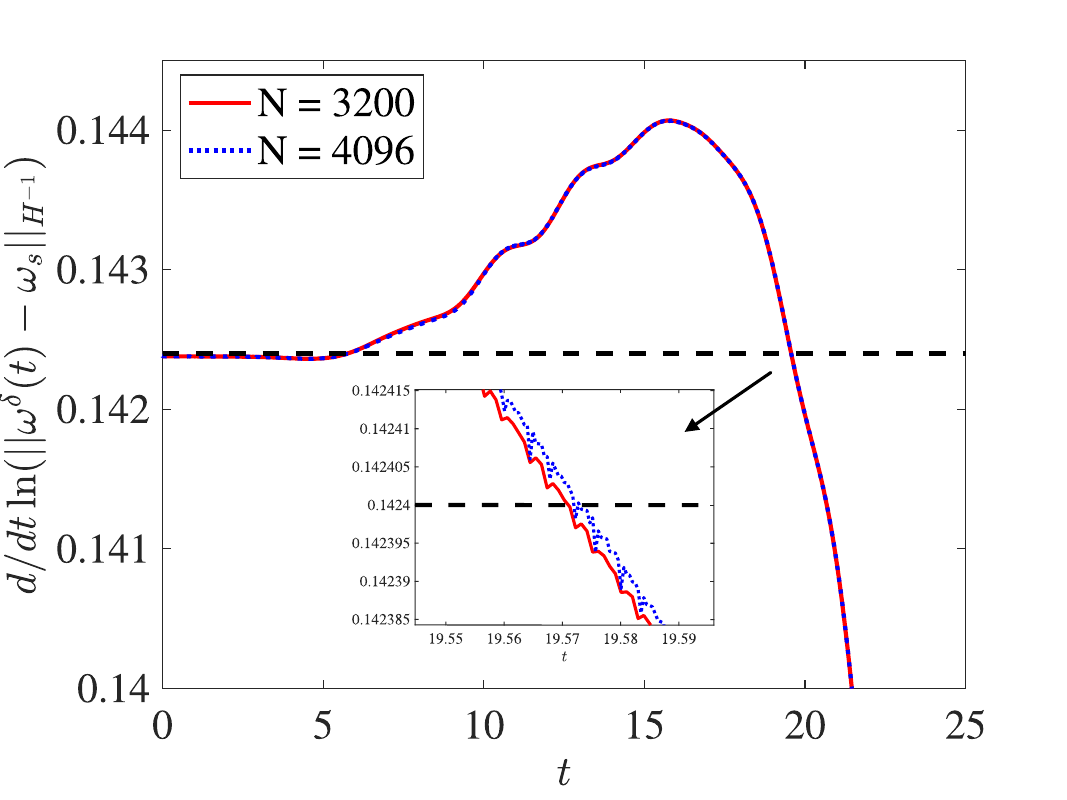}
    \caption{}
  \end{subfigure}
  \caption{{The time-dependence of (a) the norm {$\|
        \omega^{\delta}(t) - \omega_s\|_{H^{-1}}$} and (b) the growth
      rate $(d/dt) \ln (\| \omega^{\delta}(t) - \omega_s\|_{H^{-1}})$
      in the solution of the nonlinear problem \eqref{eq:vort:euler}
      with the initial condition given in terms of the eigenfunction
      $\phi_+^{3000}$ as $\omega_0 = \omega_s + \delta \cdot
      \Re\left(\phi_+^{3000}\right)/\left\|\Re\left(\phi_+^{3000}\right)\right\|_{L^2}$
      with different indicated magnitudes $\delta$.  The dashed
      horizontal line in panel (b) corresponds to
      {$\Re\left(\lambda_+^{3000}\right)$}, cf.~figure
      \ref{fig:evals}b.  In panel (c), we compare the growth rate of
      the solutions of the nonlinear problem
        \eqref{eq:vort:euler} corresponding to the same initial
      condition $\omega_0$ with $\delta = 10^{-2}$ computed using
      different indicated numerical resolutions.}}
  \label{fig:nonlinear}
\end{figure}

Finally, we consider the question about the nonlinear stability of the
equilibrium \eqref{eq:TG}. {Part (ii) of Theorem \ref{thm:Lin}
  asserts that if the eigenfunction $\phi_+$ is at least in
  $L^2(\TT^2)$, then the equilibrium is also nonlinearly unstable.  We
  emphasize that this is not a trivial statement since for
  infinite-dimensional problems such as \eqref{eq:vort:euler} a linear
  instability need not imply a nonlinear instability. In \S\,
  \ref{sec:modal_growth} we provided numerical evidence that the
  eigenfunction $\phi_+ \in H^{0.28}_0(\TT^2) \subset L^2(\TT^2)$.
  Therefore, a nonlinear instability is indeed expected and here we
  illustrate this behavior.}  In figures \ref{fig:nonlinear}a and
\ref{fig:nonlinear}b we show the time dependence of the kinetic energy
of the perturbation given by the norm $\left\| \omega^{\delta}(t) -
  \omega_s \right\|_{H^{-1}}$ and of the corresponding rate of growth
when the evolution is governed by the nonlinear problem
\eqref{eq:vort:euler} with the initial condition $\omega_0$ given in
terms of the eigenfunction $\phi_+^{3000}$ as
\begin{equation}\label{eq:omega:ptb}
\omega_0 = \omega_s + \delta \cdot
\Re\left(\phi_+^{3000}\right)/\left\|\Re\left(\phi_+^{3000}\right)\right\|_{L^2}
\end{equation}
with different indicated magnitudes $\delta$. {The time evolution
  is computed using the spatial resolution $N^2 = 3200^2$ and the time
  step $\Delta t = 2^{-10}$.}  We see that in each case the vorticity
perturbation $(\omega^{\delta}(t) - \omega_s)$ at first grows
exponentially, as is the case in the linear problem
\eqref{eq:vort:linear}, cf.~figures \ref{fig:growth0}a,b, until this
growth saturates due to nonlinear effects. Since this behavior occurs
no matter how small the norm of the initial perturbation is, the
results presented in figures \ref{fig:nonlinear}a and
\ref{fig:nonlinear}b {confirm} that equilibrium \eqref{eq:TG} is
also nonlinearly unstable. {We also perform a resolution
  refinement study to investigate whether the saturation
 evident in figures \ref{fig:nonlinear}a and
  \ref{fig:nonlinear}b is a result of an insufficient
  numerical resolution. Specifically, we compute the solution of the
  nonlinear problem \eqref{eq:vort:euler} using the same initial
  condition given by \eqref{eq:omega:ptb} with $\delta = 10^{-2}$ and
  a higher spatial resolution $N^2 = 4096^2$. As is shown in figure
  \ref{fig:nonlinear}c, the two solutions computed using different
  resolutions ($N^2 = 3200^2$ and $4096^2$) reach the nonlinear stage
  at essentially the same time and the difference in the
  growth rate of the norm is negligible. This confirms that the
  saturation seen in figures \ref{fig:nonlinear}a and
  \ref{fig:nonlinear}b is physical and not due to numerical artifacts.
  Moreover, since the kinetic energy is conserved in the Euler system
  \eqref{eq:vort:euler}, we have
\begin{equation}
\left\| \omega^{\delta}(t) - \omega_s \right\|_{H^{-1}} 
\leq\left\| \omega^{\delta}(t)\right\|_{H^{-1}} + \left\| \omega_s\right\|_{H^{-1}} 
 = \left\| \omega^{\delta}(0)\right\|_{H^{-1}} + \left\| \omega_s\right\|_{H^{-1}}.
\end{equation}
{As} the right-hand side of the above equation does not depend on
time, we know that $\left\| \omega^{\delta}(t) - \omega_s
\right\|_{H^{-1}}$ cannot grow exponentially for all times.}

\section{Summary and Conclusions}
\label{sec:final}

In this study we have considered the stability of the Taylor-Green
vortex in inviscid planar flows governed by the 2D Euler system
\eqref{eq:euler}. The Taylor-Green vortex \eqref{eq:TG} is a simple
equilibrium solution of that system characterized by a cellular
structure with hyperbolic stagnation points. In contrast to most
earlier studies
\citep{SippJacquin1998,LeblancGodeferd1999,GauHattori2014,Suzuki2018,HattoriHirota2023},
we have considered the problem in the inviscid setting where there are
important differences with respect to the viscous problem, in
particular, as regards the structure of the spectrum of the linearized
operator.  As the most important result, we have presented numerical
evidence for the presence of two distinct mechanisms of linear
instability in this flow.

First, by numerically solving the eigenvalue problem \eqref{eq:sig0}
and then integrating the linearized Euler system
\eqref{eq:vort:linear} in time, we showed the existence of an unstable
eigenvalue {$\lambda_+ \approx 0.1424 + 0.5875i$}. {Through
  a careful analysis of the behavior of the numerical approximations
  of the corresponding eigenfunction $\phi_+$ {both in the
    physical and in the Fourier space, we provided convincing evidence
    that this eigenfunction belongs to $H^{0.28}_0(\mbb T^2)$, which
    is in close agreement with the assertion in part (i) of Theorem
    \ref{thm:Lin} \citep{Lin2004}, thereby demonstrating the sharpness
    of this result. Moreover, the eigenfunction is discontinuous at
    the hyperbolic stagnation points $\x_s$.  We also showed that, in
    agreement with part (ii) of Theorem \ref{thm:Lin}, this
    eigenfunction also gives rise to a nonlinear
    instability. {In this context we note that, employing the
      complementary (with respect to the one used in
      \S\,\ref{sec:modal_growth}) form of the Sobolev embedding
      $W^{1,p^*} \hookrightarrow L^q$ \citep{af05}, where
      $1/p^* - 1/2 = 1/q$, we deduce that at the same time
      $\phi_+ \in L^q(\TT)$ with $q = 2.76$. Consequently, the initial
      condition for the 2D Euler system \eqref{eq:vort:euler} given in
      \eqref{eq:omega:ptb} in terms of this eigenfunction does not
      belong to the Yudovich class
      $L^1(\TT^2) \bigcap L^\infty(\TT^2)$ \citep{Yudovich1963} and
      therefore uniqueness of the solution cannot in general be
      guaranteed. In fact, as argued by
      \citet{Vishik2018a,Vishik2018b,BressanShen2021,Brue2024flexibilitytwodimensionaleulerflows},
      initial data in $L^q$ with $q < \infty$ could lead to nonunique
      solutions; moreover, such solutions could also exhibit anomalous
      dissipation.}  Similar properties resulting from an interplay
    between the point spectrum and the essential spectrum of the
    linearized Euler operator were recently revealed in the stability
    analysis of the Lamb-Chaplygin dipole by} \citet{Protas2024}.}

Second, we illustrated a nonmodal mechanism of instability growth
which involves a {\em continuous} family of uncorrelated functions,
rather than a single eigenfunction of the linearized operator $\L$.
{This nonmodal instability is tied to perturbations characterized
  by highly localized oscillatory features, a mechanism that has also
  been studied by \cite{Sengupta:2011, Sengupta:2020} who showed that
  the corresponding component in the energy spectrum plays an
  important role in the transition to turbulence in wall-bounded
  flows.}  Unlike the eigenfunction $\phi_+$, the optimal initial
conditions $\wopt_0^N$ depend on the function space in which they are
defined and we considered two Sobolev spaces, namely, $H^1(\TT^2)$ and
$H^{-1}(\TT^2)$. Constructed by solving a suitable PDE optimization
problem, Problem \ref{prob:ess:growth}, the resulting flows saturate
the estimates on the growth of the semigroup $e^{t \L}$ implied by the
essential spectrum $\sigess(\L)$ via the Spectral Mapping Theorem
\eqref{eq:SMT} as the numerical resolution is refined.  Using some
generic vorticity field as the initial condition $w_0$ in the linear
problem \eqref{eq:vort:linear} will result, after some transient, in
the corresponding solution {growing} as $\O\left( e^{\Re(\lambda_+) t}
\right)$. {This points} to the absence of eigenvalues {\em outside}
the essential spectrum $\sigess(\L)$.

The optimal initial conditions obtained by solving Problem
\ref{prob:ess:growth} exhibit a similar spatial structure to the
initial conditions found by \citet{GauHattori2014} which were obtained
by maximizing a weighted norm $||e^{t\L} w_0||_{H^{2}}$ for different
$t$ in a viscous flow. However, the key difference is that the largest
growth rate found in that study was essentially equal to the real part
of the most unstable eigenvalue, meaning that in that case this growth
was effectively realized by the most unstable eigenmode. In contrast,
in the present study the largest growth rate is in fact larger than
$\Re(\lambda_+)$ and this behavior is not realized by an
eigenfunction, but by a continuous family of uncorrelated
distributions.  
We emphasize that this mechanism is intrinsically
linked to the inviscid and infinite-dimensional nature of the
operator $\L$ and, as such, is fundamentally different from the
transient growth of perturbations arising as a result of the
nonnormality of the eigenvectors of a linear operator
\citep{Schmid2001book}. These results are consistent with the
predictions of the WKB analysis, which points to linear instabilities
growing as $\O(e^{\mumax t})$ \citep{FriedlanderVishik1991,lh91}.
However, in our study we are also able to characterize the global
spatial structure of this instability paying attention to the
regularity of the perturbations, which is beyond reach of the WKB
analysis.

We remark that the solutions discussed here are not smooth functions
of the space variable $\x$ {and exhibit} small-scale features
localized near the hyperbolic stagnation points $\x_s$. Therefore,
they cannot be fully resolved in numerical computations with any {\em
  finite} resolution $N^2$, cf.~\S\,\ref{sec:numer}. However, a
Galerkin truncation such as \eqref{eq:fhat} used to construct
numerical solutions, together with the resolution-dependent low-pass
filter employed in the solution of the time-dependent problems
\eqref{eq:vort:euler} and \eqref{eq:vort:linear}, can be viewed as a
regularization of the original system with the effect {decreasing} as
the spatial resolution {$N^2$} is refined. The key quantities
characterizing the instability, namely, the eigenvalues $\lambda_+^N$
and the corresponding eigenfunctions $\phi_+^N$, as well as the growth
rates {$(d/dt) \ln (\|w^N(t)\|_{L^2})$,
  $(d/dt) \ln (\|w^N(t)\|_{H^{1}})$, and
  $(d/dt) \ln (\|w^N(t)\|_{H^{-1}})$,} are shown to converge to
well-defined limits as the numerical resolution $N^2$ is refined,
cf.~figures {\ref{fig:evals}b, \ref{fig:eigfun}c--e,
  \ref{fig:growth0}a,b,} \ref{fig:s:1}a and \ref{fig:s:-1}a.

\begin{figure}
  \centering
  \begin{subfigure}{0.48\textwidth}
\includegraphics[width=\textwidth]{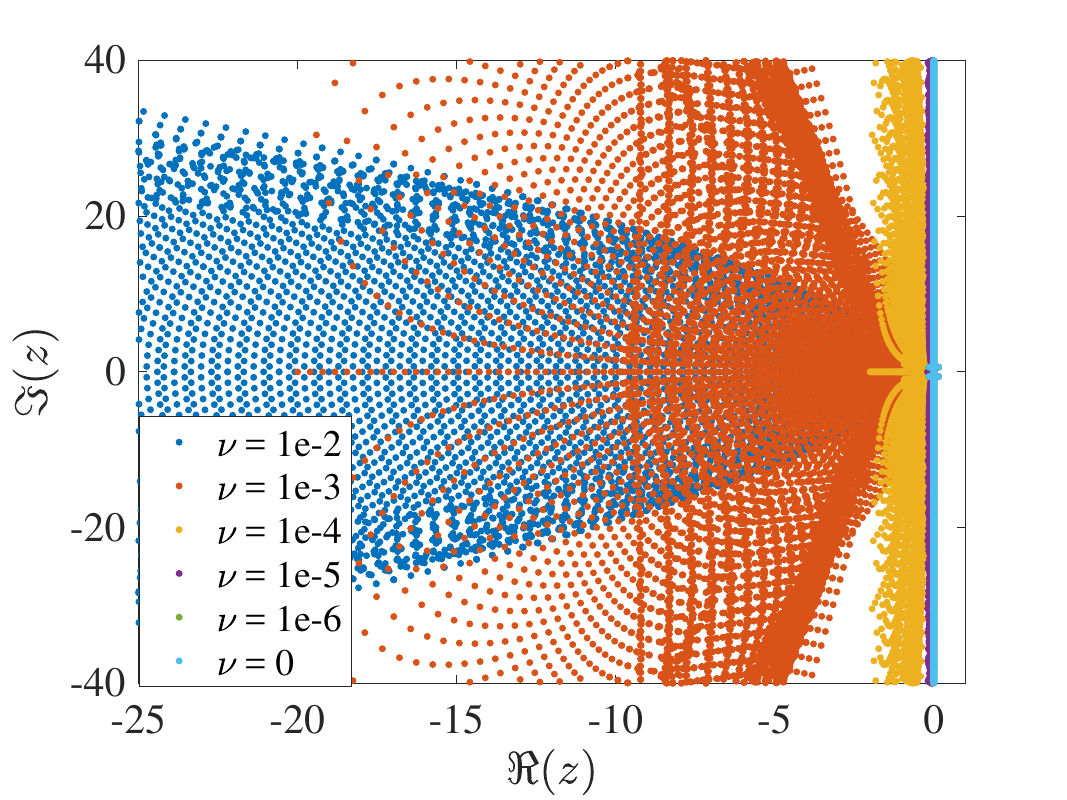}
\caption{}
\end{subfigure}
\hfill
 \begin{subfigure}{0.48\textwidth}
\includegraphics[width=\textwidth]{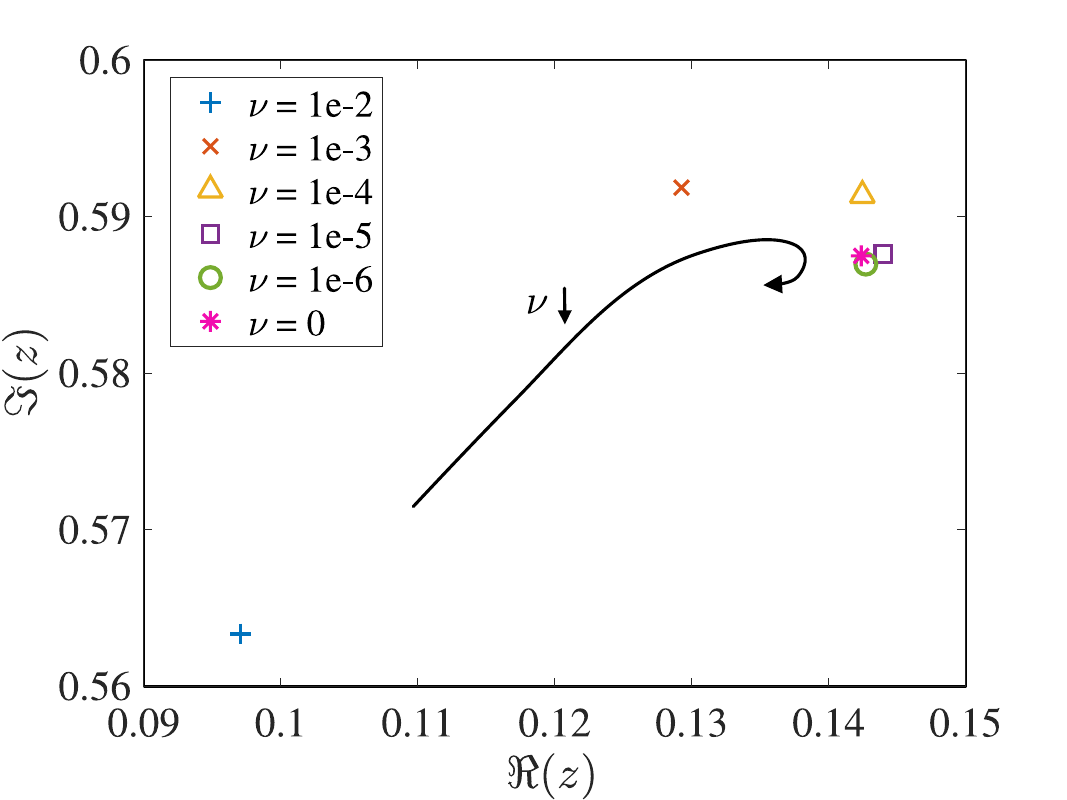}
\caption{}
\end{subfigure}
\caption{(a) Eigenvalues of the operator $\L +\nu \Delta$ for
    different indicated values of $\nu$; they are obtained by solving
    the discrete eigenvalue problem \eqref{eq:evpN} modified to
    include the dissipative term $\nu \Delta$ and using the resolution
    $N^2 = 200^2$.  (b) Magnification of the neighbourhood of the
  eigenvalue $\lambda_+^{3000}$ with the arrow indicating the
  trend with the decrease of $\nu$; the eigenvalues
    shown in this panel are computed using the Krylov subspace method
    and the resolution $N^2 = 3000^2$.}
  \label{fig:evalsnu}
\end{figure}

\begin{figure}
\centering
 \begin{subfigure}{0.48\textwidth}
\includegraphics[width=\textwidth]{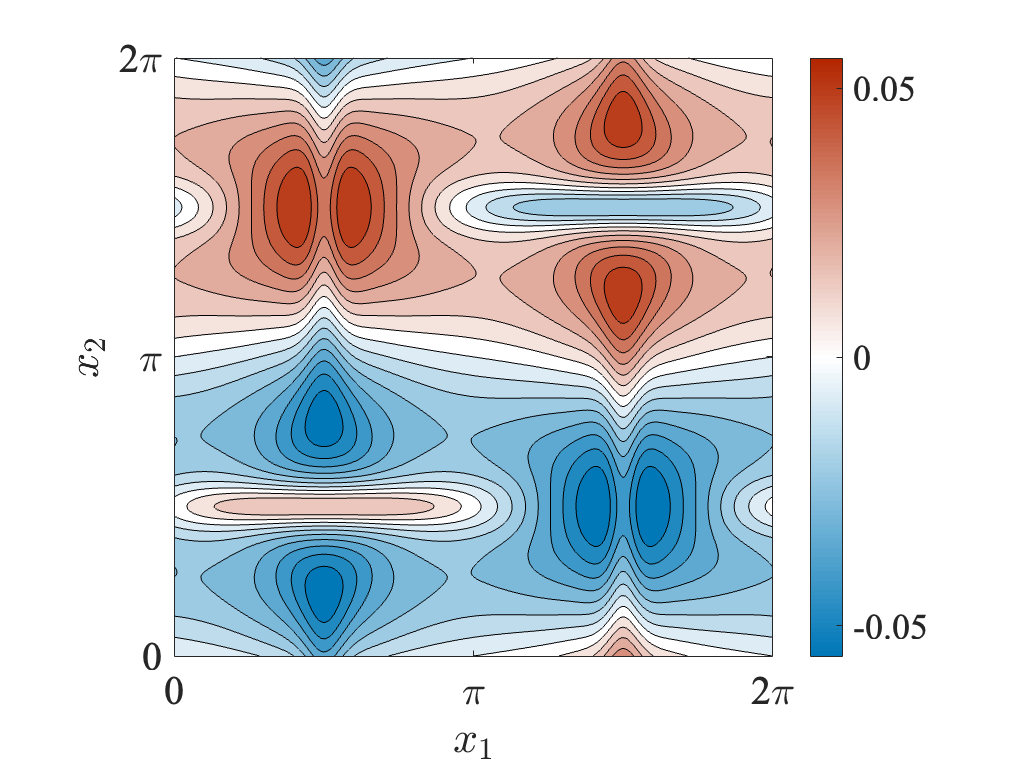}
\caption{}
\end{subfigure}
\begin{subfigure}{0.48\textwidth}
\includegraphics[width=\textwidth]{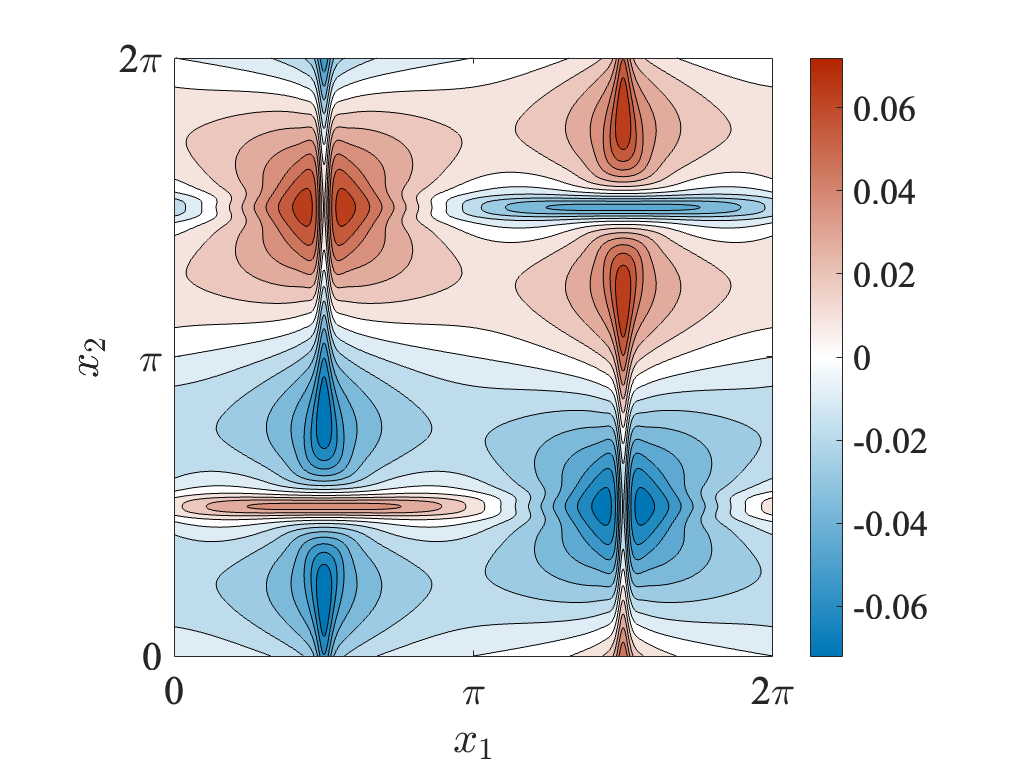}
\caption{}
\end{subfigure}

\begin{subfigure}{0.48\textwidth}
\includegraphics[width=\textwidth]{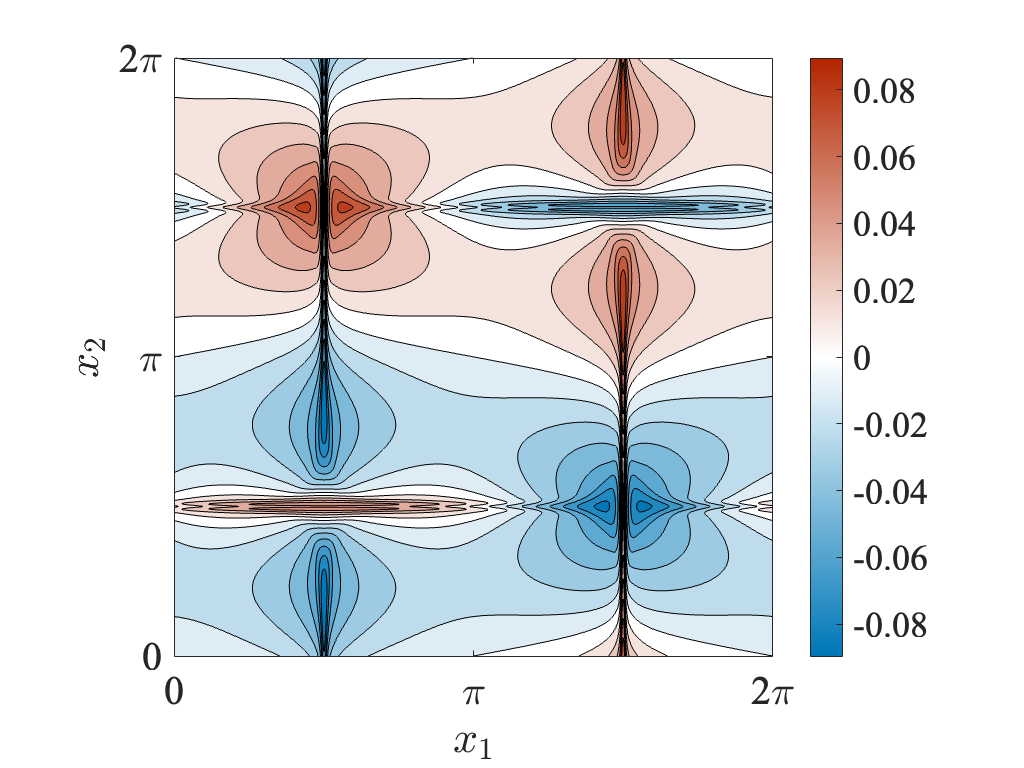}
\caption{}
\end{subfigure}
\begin{subfigure}{0.48\textwidth}
\includegraphics[width=\textwidth]{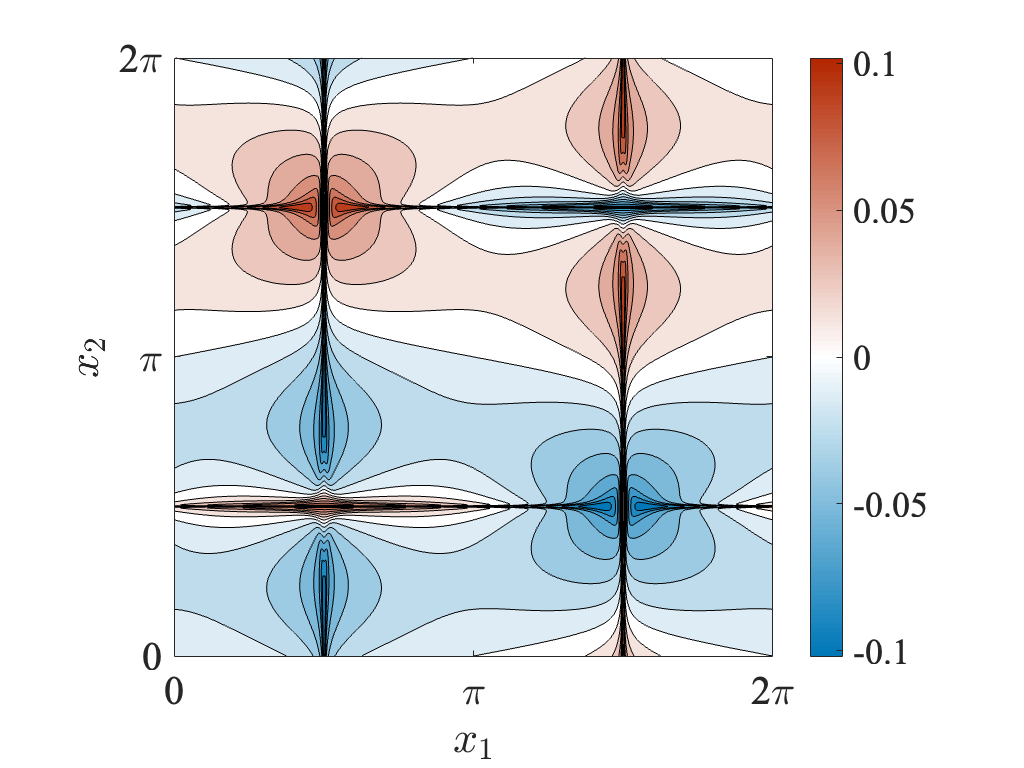}
\caption{}
\end{subfigure}

\caption{Contour plots of the real part of the eigenfunctions
  corresponding to the unstable eigenvalues of $\L +\nu \Delta$ shown
  in figure \ref{fig:evalsnu}(b): (a) $\nu = 10^{-2}$, (b)
  $\nu = 10^{-3}$, (c) $\nu = 10^{-4}$, and (d) $\nu = 10^{-5}$.}
  \label{fig:NS:eigfun}
\end{figure}

Since it is known that in the viscous case the spectrum of the
linearized Navier-Stokes operator consists of the discrete spectrum
only, it is interesting to investigate the effect of viscous
perturbations on the spectrum of the linearized Euler operator $\L$,
cf.~figure \ref{fig:evals}.  In figure \ref{fig:evalsnu}a we show
solutions of the discrete eigenvalue problem \eqref{eq:evpN} modified
to include a dissipative term proportional to the viscosity $\nu$,
i.e., for the perturbed operator $\L + \nu\Delta$ for different
indicated values of $\nu$.  We see that with the addition of viscosity
the essential spectrum in $L^2(\TT^2)$, which in the inviscid problem
coincides with the {imaginary} axis $i\RR$, disintegrates into a
number of discrete eigenvalues located inside a parabolic region in
the left half-plane $\Re(\lambda) < 0$. At the same time, as is
evident from figure \ref{fig:evalsnu}b, the discrete eigenvalue
$\lambda_+$ is perturbed, but remains on the right half-plane.  The
unstable eigenvalue obtained with $\nu = 10^{-5}$ is, after a suitable
rescaling, close to the result reported by \citet{HattoriHirota2023}
{with a $0.7\%$ relative error}.  {As is evident from figure
  \ref{fig:evalsnu}b, the unstable eigenvalues of $\L + \nu\Delta$
  converge to the unstable eigenvalues of $\L$ in the limit of
  vanishing viscosity, which is consistent with the theoretical
  results by \citet{ShvydkoyFriedlander2008}.  This further
  demonstrates that the linear instabilities considered here are
  fundamentally inviscid properties. To close this discussion, in
  figure \ref{fig:NS:eigfun}, we plot the real part of the
  eigenfunctions corresponding to the unstable eigenvalues shown in
  figure \ref{fig:evalsnu}b for $\nu = 10^{-2}, 10^{-3}, 10^{-4}$, and
  $10^{-5}$.  Similarly to the unstable eigenfunction obtained in the
  inviscid case, cf.~figure \ref{fig:eigfun}a,b, they all reveal an
  odd symmetry while the Taylor-Green vortex possesses an even
  symmetry.  As $\nu$ decreases, these eigenfunctions become
  concentrated along the heteroclinic orbits.  \cite{Sengupta:2018}
  showed that numerical errors induce a symmetry-breaking instability
  in the computation of the viscous evolution of the 2D Taylor-Green
  vortices. We think this is because the numerical errors contain
  components proportional to the odd unstable eigenfunctions.}
 
In terms of future work, a natural question to consider is an
extension of the problems studied here to the stability of 2D
Taylor-Green vortices in 3D Euler flows \citep{HattoriHirota2023}.
However, mathematically rigorous results are much more limited in 3D
due to the presence of the vortex-stretching term $(\bds u \cdot
\nabla)\bds \omega$ in the 3D Euler equations.  In addition, in 3D
$\mu_{\max}$ cannot be easily computed by evaluating $\nabla \bds u$
at hyperbolic stagnation points and a counterpart of the Spectral
Mapping Theorem \eqref{eq:SMT} is not available. {Furthermore,
  the eigenfunction shown in figure \ref{fig:eigfun} and the optimal
  initial conditions shown in figures \ref{fig:cont:s:1} and
  \ref{fig:cont:s:-1} reveal the absence of a smallest scale.
  Therefore, in order to resolve the small-scale features dominating
  these objects in a computationally efficient manner, in the future
  we plan to use discretization techniques combining nonuniform grids
  \citep{Sengupta:2018} with adaptive mesh refinement
  \citep{Ceniceros:Hou:2001, Di:2008:adaptive}.}
{Finally, it is an interesting question whether the
  nonmodal growth discussed in \S\,\ref{sec:nonmodal_growth}
 can also reach the nonlinear stage and lead to turbulence.
  However, to answer this question, one needs to solve Problem
  \ref{prob:ess:growth} over a much longer time interval and with a
  much higher numerical resolution, which would necessitate
    larger computational resources than currently available. One
  potential future direction to address this question is to find ways
  to reduce the dimension of our search space by using a priori
  knowledge about the optimal initial conditions. }

\section*{Acknowledgements}

The authors are thankful to Matthew Colbrook {and Hao Jia} for
discussions about the numerical solution of eigenvalue problems for
non-self-adjoint infinite-dimensional operators. XZ and BP acknowledge
the support through an NSERC (Canada) Discovery Grant. RS was
partially supported through an NSF grant DMS--2107956 and
DMS--2405326. {BP and RS would like to thank the Isaac Newton
  Institute for Mathematical Sciences for support and hospitality
  during the programme ``Mathematical aspects of turbulence: where do
  we stand?'' where this research was initiated. This work was
  supported by EPSRC grant number EP/R014604/1. } Computational
resources were provided by the Digital Research Alliance of Canada
under its Resource Allocation Competition.


\FloatBarrier
\appendix
\section{Riemannian Conjugate Gradient Approach}
\label{sec:RCG}

Problems \ref{prob:ess:growth} is solved numerically with a Riemannian
conjugate gradient method \citep{DanailaProtas2017}.  {At
 each ($n$-th) iteration, the method consists of three steps.
  First, we project the gradient $\nabla J\left(w_0^{(n)}\right)$
  given in \eqref{eq:gradJ2} onto the space tangent to $\M$ at
  $w_0^{(n)}$. Then, we use the previous search direction, denoted
  $d^{(n-1)}$, to construct a Riemannian conjugate ascent direction
  using a suitable vector transport operation, and combine it with the
  projected gradient obtained in the first step to construct the
  current search direction $d^{(n)}$. Finally, we retract the
  resulting state back to the constraint manifold $\M$. } A local
maximizer of Problem \ref{prob:ess:growth} is obtained as $\wopt_0 =
\lim_{n \rightarrow \infty} {w_0}^{(n)}$, where the successive
approximations ${w^{(n)}_0}$ are therefore determined with the
iterative formula
\begin{equation} \label{eq:RCG}
{w^{(n+1)}_0  = \PR\Big[w^{(n)} _0+ \tau_n d^{(n)}\Big], \qquad n=0,1,\dots,}
\end{equation}
{where ${w^{(0)}_0}$ is the initial guess.} Here $\PR: {H_0^m(\TT^2)} \to \mc M$ is the retraction operator defined by
\citep{ams08}
\begin{equation}
\PR (w) := \frac{w}{||w||_{H^m}}, \qquad w \in {H^m_0(\TT^2)}
\end{equation}
{which normalizes the state ${w^{(n)}_0 + \tau_n d^{(n)}}$
  to pull it} back to the constraint manifold $\mc M$ defined
in Problem \eqref{prob:ess:growth}.  {The optimal step size
  $\tau_n$ is obtained by solving an arc-search problem to find the
  step size $\tau$ such that the objective functional $J$ achieves its
  maximum along the curve ${\{R[w^{(n)}_0 + \tau d^{(n)}], \; \tau >
    0\}}$ on the manifold $\mc M$,} i.e.,
\begin{equation} \label{eq:arcsearch}
{\tau_n = \argmax_{\tau > 0} \left\{J
\left(\PR\Big[w^{(n)}_0 + \tau d^{(n)}\Big]\right)  \right\}.}
\end{equation}
{This problem is} solved with a suitable derivative-free approach, such as a
variant of Brent's algorithm ~\citep{pftv86}.
\begin{figure}
  \centering
  \includegraphics[width=0.6\textwidth]{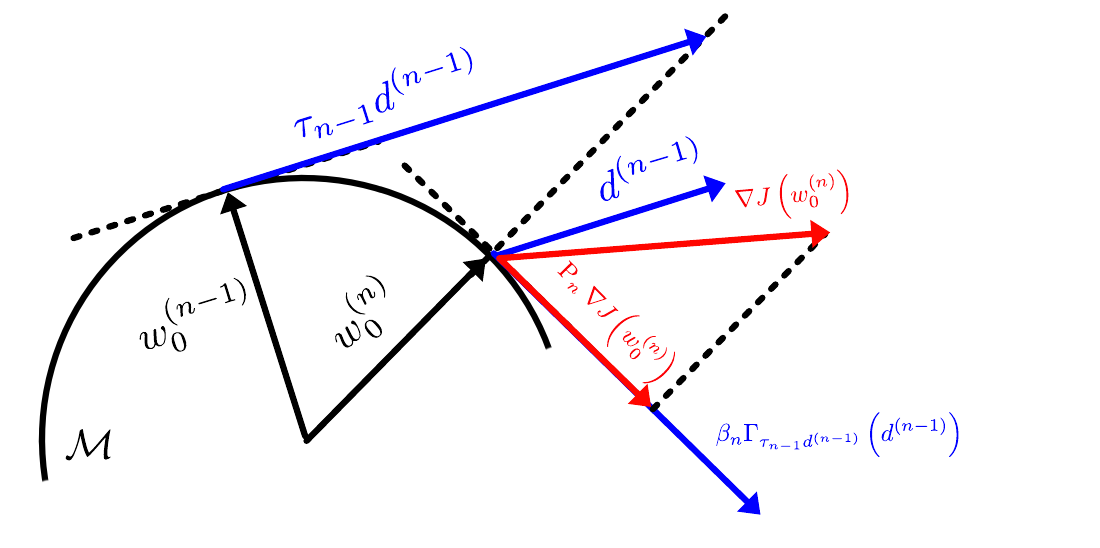}
  \caption{Schematic illustration of the Riemannian
      conjugate gradient method \eqref{eq:RCG}.}
  \label{fig:RCG}
\end{figure}

{We denote the space tangent to the manifold $\M$ at
  $w_0^{(n)}$ by $\mc T_{w_0^{(n)}} \M$.} The search direction
$d^{(n)}$ in \eqref{eq:RCG} {belongs to $\mc T_{w_0^{(n)}} \M$
  and } is computed as
\begin{equation}\label{eq:search:conjg}
\begin{aligned}
d^{(0)} &= \PP_0 \nabla J\left(w^{(0)}_0\right),\\
d^{(n)} &= \PP_n  \nabla J\left(w^{(n)}_0\right)+ \beta_n \Gamma_{\tau_{n-1} d^{(n-1)}}\left(d^{(n-1)}\right),
\qquad n\geq 1.
\end{aligned}
\end{equation}
{As is illustrated in figure \ref{fig:RCG}, the
  projection operator $\PP_n \: : \; {H^m_0(\TT^2)} \rightarrow \mc
  T_{w_0^{(n)}} \mc M$ realizes an orthogonal projection onto the
  linear subspace $\mc T_{w_0^{(n)}} \mc M$.}  It is defined by the
relation
\begin{equation}\label{eq:proj}
\PP_n w := w - \la w, \; \nu \ra_{H^m} \nu, \qquad 
\nu = {\frac{w_0^{(n)}}{\left|\left| w_0^{(n)} \right|\right|_{H^m}}.}
\end{equation}
Since ${\PP_n \nabla J\left(w^{(n)}_0\right) \in \mc T_{w^{(n)}} \mc M}$
whereas $d^{(n-1)} \in \mc T_{w^{(n-1)}} \mc M$, these two elements
belong to different linear spaces and as such cannot be directly
added. Therefore, we utilize the vector transport $\Gamma$
defined in terms of the differentiated retraction
\citep{ams08} to map the element $d^{(n-1)}$ from the subspace
{$\mc T_{w_0^{(n-1)}} \mc M$ to $\mc T_{w_0^{(n)}}\mc M$.}  For any
$w \in \mc M$ and $ \xi_w, \varphi_w \in \mc T_w \mc M$, we define
\begin{equation}\label{eq:vec:transpt}
\begin{aligned}
  \Gamma_{\varphi_{w}}(\xi_{w}) &:= \frac{d}{ds} \PR(w + {\varphi_{w} + s \xi_{w}}) \big|_{s=0}  \\
  &= \frac{1}{\| w + \varphi_{w} \|_{H^m}}\left[ \xi_{w} - \frac{\langle w+ \varphi_{w}, \; \xi_{w} \rangle_{H^m}}{\| w + \varphi_{w} \|_{H^m}^2} (w+ \varphi_w) \right].
  \end{aligned}
 \end{equation}
 Setting $w = w^{(n)}$, $\varphi_w = \tau_{n-1} d^{(n-1)}$, and $\xi_w
 = d^{(n-1)}$, we then obtain
\begin{equation}
{
\Gamma_{\tau_{n-1} d^{(n-1)}}\left(d^{(n-1)}\right) = 
\frac{1}{\| w^{(n)}_0\|_{H^m}}\PP_n d^{(n-1)}.}
\end{equation}
A schematic illustration of the Riemannian conjugate gradient
  method \eqref{eq:RCG} is shown in figure \ref{fig:RCG}.

The ``momentum'' term $\beta_n$ in (\ref{eq:search:conjg}) is
chosen to enforce the conjugacy of consecutive search directions and
is computed using the Polak-Ribi\`ere approach \citep{nw00}
\begin{equation}
{
\beta_n = \frac{\Big\la \PP_n  \nabla J\left(w^{(n)}_0\right), 
\Big( \PP_n \nabla J\left(w^{(n)}_0\right) -
\Gamma_{\tau_{n-1} d^{n-1}}\PP_{n-1}  \nabla J\left(w^{(n-1)}_0\right) \Big)\Big\ra_{H^m}}
{\Big|\Big|  \PP_{n-1}  \nabla J\left(w^{(n-1)}_0\right)\Big|\Big|^2_{H^m}}.
}
\end{equation}
In our computation, we restart algorithm \eqref{eq:RCG} by setting
$\beta_n = 0$ based on the following two criteria necessary from both
the theoretical and practical point of view as they help erase
obsolete information from earlier iterations \citep{nw00}
\begin{itemize}
\item[(1)]\; $n = 20 k$, $k \in \mbb Z_+$,
\smallskip
\item[(2)]\;  The search direction $d^{(n)}$ fails to be an ascent direction, i.e., 
\begin{equation}
{
\frac{\Big\la d^{(n)}, \PP_n \nabla J\left(w^{(n)}_0\right) \Big\ra_{H^m}}
{\Big|\Big| d^{(n)}\Big|\Big|_{H^m}\Big|\Big| \PP_n \nabla J\left(w^{(n)}_0\right)\Big|\Big|_{H^m}}
{< \tol1}, \qquad 0 < \tol1 \ll1.
}
\end{equation}

\end{itemize}

\begin{figure}
  \centering
\includegraphics[width=0.6\textwidth]{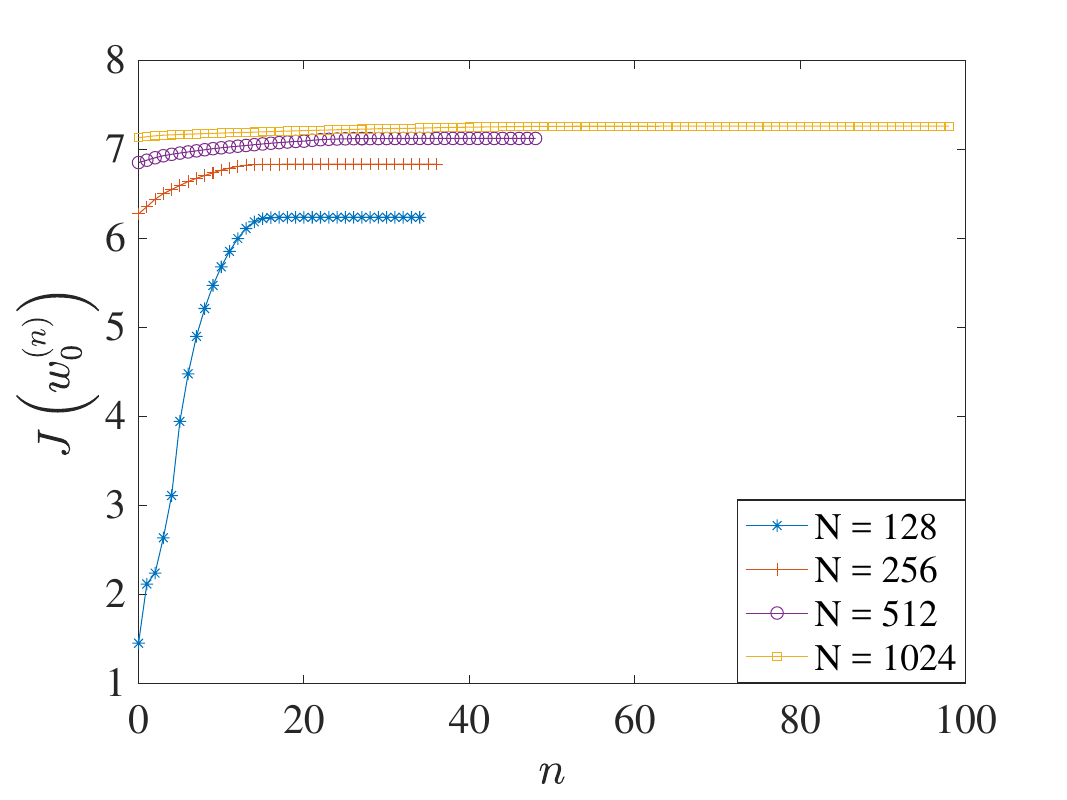}
\caption{Dependence of the objective functional $J\left(w_0^{(n)}\right)$ on the iteration index $n$ for different numerical resolutions.}
  \label{fig:iteration}
\end{figure}

Iterations \eqref{eq:RCG} are declared converged when the relative
change of the objective functional \eqref{eq:J0} between two
consecutive iterations becomes smaller than a specified tolerance
$0 < \tol2 \ll 1$, i.e., when
\begin{equation}
{
0 \leq \frac{J\left(w^{(n+1)}_0\right) - J\left(w^{(n)}_0\right)}{J \left(w^{(n)}_0\right)}
< {\tol2}.
}
\end{equation}
{In practice, we set $\tol1 = \tol2 = 10^{-10}$. In order
    to illustrate the performance of algorithm \eqref{eq:RCG}, figure
  \ref{fig:iteration} shows the values of the objective functional
  $J\left(w_0^{(n)}\right)$ for $m = 1$ as function of the iteration
  index $n$.  As explained in \S\,\ref{sec:nonmodal_growth}, we solve
  Problem~\ref{prob:ess:growth} using increasing resolutions $N^2 =
  128^2, 256^2, 512^2, 1024^2$, and use the optimal initial condition
  $\wopt_0^N$ obtained with the resolution $N^2$ as the initial guess
  in the iteration \eqref{eq:RCG} with the resolution
  $(2N)^2$.  As shown in the figure, the method requires more
  iterations to converge for higher resolutions due to an
    increased number of degrees of freedom.}

\FloatBarrier


\bigskip
\noindent
{\bf Declaration of Interests.} The authors report no conflict of interest.




\end{document}